\documentclass[aps,pre,twocolumn,longbibliography]{revtex4-2}

\usepackage{mwe}
\usepackage{graphicx}
\usepackage{dcolumn}
\usepackage{bm}
\usepackage{hyperref}
\usepackage{mathtools}

\usepackage{xcolor}
\usepackage{amsmath}
\usepackage{amssymb}

\usepackage{titletoc}
\usepackage{titlesec}

\usepackage{newfloat}
\DeclareFloatingEnvironment[name={Supplementary Figure}]{suppfigure}

\newcommand{\ds}{\displaystyle}

\begin{document}

\title{Spreading dynamics in networks under context-dependent behavior}

\author{Giulio Burgio}
\author{Sergio G\'omez}
\author{Alex Arenas}

\email{alexandre.arenas@urv.cat}
\affiliation{
Departament d'Enginyeria Inform\`atica i Matem\`atiques, Universitat Rovira i Virgili, 43007 Tarragona, Spain
}

\begin{abstract}

In some systems, the behavior of the constituent units can create a \lq context\rq\ that modifies the direct interactions among them. This mechanism of indirect modification inspired us to develop a minimal model of context-dependent spreading. In our model, agents actively impede (favor) or not diffusion during an interaction, depending on the behavior they observe among all the peers in the group within which that interaction occurs. We divide the population into two behavioral types and provide a mean-field theory to parametrize mixing patterns of arbitrary type-assortativity within groups of any size. As an application, we examine an epidemic spreading model with context-dependent adoption of prophylactic tools such as face-masks. By analyzing the distributions of groups' size and type-composition, we uncover a rich phenomenology for the basic reproduction number and the endemic state. We analytically show how changing the group organization of contacts can either facilitate or hinder epidemic spreading, eventually moving the system from the subcritical to the supercritical phase and vice versa, depending mainly on sociological factors, such as whether the prophylactic behavior is hardly or easily induced. More generally, our work provides a theoretical foundation to model higher-order contexts and analyze their dynamical implications, envisioning a broad theory of context-dependent interactions that would allow for a new systematic investigation of a variety of complex systems.

\end{abstract}

\maketitle

\section{INTRODUCTION}
\label{sec:intro}

Real systems often exhibit higher-order, group interactions that cannot be reduced to combinations of pairwise interactions~\cite{Gleeson_2013,battiston2020networks,bick2021higher}. Even when an interaction is pairwise e.g., disease spreading, if the behavior of the involved units is modified by the co-presence of other units (e.g., other people), the interaction cannot be considered in isolation. The group as a whole defines a ``context'' which alters the (direct) interactions taking place within it. These higher-order interactions are known as ``indirect modifications'' in ecology~\cite{wootton1993indirect, okuyama2012model,levine2017beyond}, whereby the functional relationship between two species is altered by the presence of a third, trophically disconnected species.

In opinion formation, for instance, the direct influence that an individual can exert on another can be modified by the opinions~\cite{sahasrabuddhe2021modelling} or the individual characteristics~\cite{glaeser2014does} of the others in the group. In linguistics, when an individual moves to a new region, the learning rate of the local language depends on how often the locals use that language in the presence of that individual~\cite{sole2010diversity}. They could bear the cost of switching to another language (e.g., English) for a more inclusive conversation, but eventually slowing down the diffusion of the local one; or could instead stick to it, enforcing its diffusion.

In infectious spreading, recent studies have emphasized the importance of considering the group organization of contacts in understanding and controlling the contagion process~\cite{bodo2016sis,st2021master,st2021social,st2021universal}. In this regard, the adoption of nonpharmaceutical prophylactic behaviors, such as face-mask wearing and physical distancing, proves to be a viable and effective strategy for epidemic control~\cite{lai2012effectiveness,eikenberry2020mask,worby2020face,catching2021examining}.
Existing studies on adaptive network models for the spread of contagious disease have considered agents whose behavior depends on individual characteristics, on the observed behavior among peers, and on external sources of information, within both well-mixed and structured populations (see references [62-89] in the review by Benson et al.~\cite{bedson2021review}, and the review by Wang et al.~\cite{wang2015coupled}). However, these studies do not distinguish between potentially infectious contacts that occur in isolation versus those that occur in the co-presence of other individuals. This distinction becomes essential when accounting for behavioral adoption, and our proposed model addresses this issue by introducing higher-order context dependence. This framework provides several benefits over existing models, including greater theoretical flexibility, the ability to gain deeper theoretical insights, and the ability to infer parameter values from real-world data. Moreover, our model allows for the analysis of the effects of changing the frequency and assortativity at each group size, which cannot be captured by simpler adaptive models based on pairwise interactions. Indeed, even though each transmission concerns only two people, the likelihood of transmission is indirectly affected by the way contacts are organized within groups, for the adoption of the prophylactic behavior (e.g., wearing a face-mask) depends on the level of adoption an individual observes in the entire group~\cite{asch1955opinions,bavel2020using,bokemper2021beliefs}. Adoption is thus mutable, contextual: an individual may exhibit one behavior or another depending on the current context.

This implies that an appropriate description of the system needs to take into account either that contacts generally occur within groups, and that, concurrently to the infectious spreading, there is a decision-making process at the group level that shapes the behavior of the individuals therein. Current models either consider the group structure~\cite{bodo2016sis,st2021master,st2021social,st2021universal} or the disease-behavior coupling~\cite{bedson2021review}, but never both. In this work, we fill this gap presenting a general model of context-dependent spreading. This is representative of an entire class of models, each identified by the specific spreading process considered and by how agents' behavior is affected by the context. A degenerate subclass consists of models where behavior is context-independent, such as epidemic models where individuals can adopt permanent prophylactic tools to avoid transmission (e.g., vaccines), or information spreading models where agents divide into active sources or spreaders and passive consumers.

To describe context-dependent behavior we divide individuals into behavioral types, distinguished by a different propensity to adopt a certain behavior. The correlation between individual characteristics and behavior has been observed in various forms, among which in the adoption of prophylactic measures~\cite{Centola2010spread,Centola2011,barcelo2020voluntary}. At the same time, similar individuals not only behave similarly, they are also more likely to interact among them than with others, a phenomenon known in sociology as homophily~\cite{McPherson2001}. The local dynamics ---and the global one emerging from it--- is thus determined by both, the behavior of the types and the way they mix~\cite{burgio2021homophily,burgio2022homophily,watanabe2022impact,rizi2022epidemic,hiraoka2022herd}. To account for this, we first provide in Sec.~\ref{sec:modeling} a mean-field approximation to the higher-order structure of interactions in a population divided into two types. We thus formulate an annealed heterogeneous mixing in hypergraphs, which is a contribution in itself. Then, in Sec.~\ref{sec:behavioral}, we model the context-dependency of agents' behavior. Together, Secs.~\ref{sec:modeling} and \ref{sec:behavioral} provide the theoretical foundation for the study of context-dependent processes analytically. As an application of our theory, in Sec~\ref{sec:spreading} we consider an epidemic spreading model incorporating context-dependent adoption of prophylactic behavior, then explored in Sec.~\ref{sec:results}. There, we reveal a rich phenomenology for the basic reproduction number and the endemic state in relation to the size and type-composition distributions of the groups, unveiling the important dynamical implications of accounting for the indirect modifications of the pairwise contagion created by a higher-order context.

\section{THE MODEL}
\label{sec:the_model}

Let us consider a population of agents within which a certain entity (e.g., a pathogen, a rumour) can spread from one agent to another in a pairwise fashion. Such an interaction generally occurs in the presence of other agents not directly participating to it, i.e., within a group of some size (in particular, of size two if no other agent is present). If the co-present agents can affect the pairwise interaction, the group itself mediates an interaction: an indirect modification. Thus, we henceforth use the term  ``group'' or ``higher-order interaction'' to indicate a group of agents, defining the context within which the pairwise (direct) interactions conveying the spread take place between any two agents in the group.

Every agent, during each interaction, can either actively behave to modify (hinder or favor) the spread or not. Whether it chooses to adopt such behavior or not is determined by both the behavior it observes among the peers involved in the interaction and some intrinsic characteristics of its own. To account for the latter, suppose to partition the population into two behavioral types, labeled through letters A and N, standing for \lq adoption-inclined\rq\ and \lq not adoption-inclined\rq: accordingly, type A is assigned to those agents which are intrinsically more prone to adopt an active behavior. Then, every time a group of agents forms, each of them chooses a behavior and a spreading event is let to happen within any pair of them. Note the assumption of timescale separation we are implicitly making: the agents decide their behavior instantly, as soon as the group forms, to then let the ---slow---spreading dynamics unfold throughout the duration of the group interaction. This is a reasonable (and convenient) hypothesis in some cases, like the one considered in this work, but could be not in others. Anyway, there is no technical difficulty in relaxing it, still leaving intact the core of the model---indirect modifications generated by context-dependency.

The minimal description given so far does not yet specify how the interactions take place. Among the many possible ones, we henceforth consider the following. At time step $t$, a generic agent $i$ is involved in $k_i^{(n)}(t)$ group interactions of size $n\in\{2,\dots,n_\text{max}\}$. A group of size $n$ can be represented as a hyperedge of cardinality $n$ (or $n$-edge) of a hypergraph~\cite{bretto2013hypergraph}, whose nodes represent the agents. ``Group'' and ``hyperedge'' are synonyms here, as well as ``node'' and ``agent'', therefore we will use them indistinctly. The composition of the groups formed at time $t$ can be encoded in the adjacency tensors $\left\{{A_t}^{(n)}\right\}$, where ${A_t}^{(n)}_{i_1,\dots,i_n}=1$ if the group $\{i_1,\dots,i_n\}$ is formed at time $t$, and $0$ otherwise.

\begin{figure}
    \centering
    \includegraphics[width = 1.\linewidth]{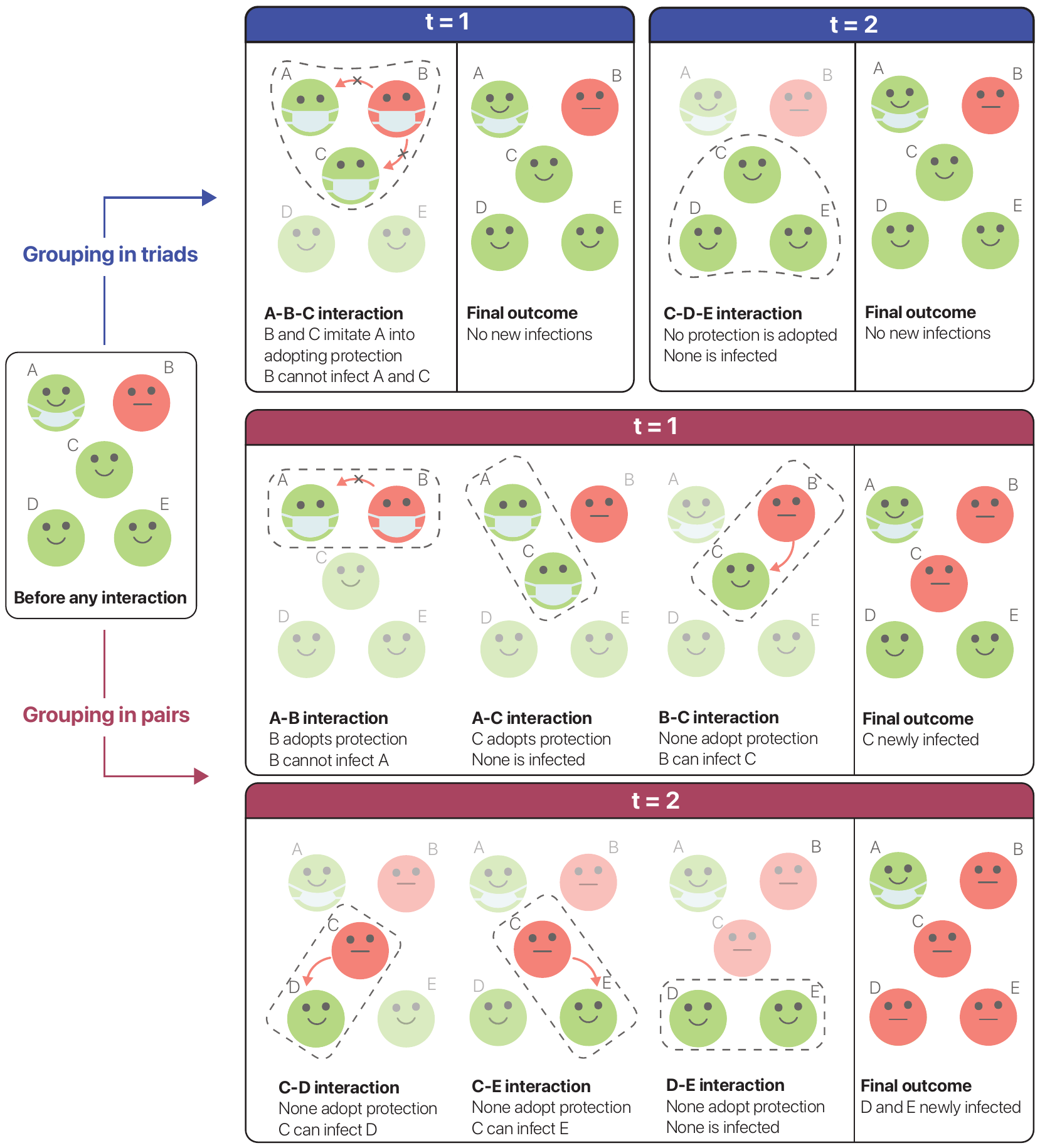}
    \caption{Example of a context-dependent spreading where the agents can act in order to avoid it (in either directions) by adopting a prophylactic behavior, represented as a worn face-mask. At each time $t$, some group interactions (dash-outlined) take place. During each interaction, an infection can occur from any infected (red) agent to any susceptible (green) one. In this example, A-type nodes (represented with a ready-to-use, lowered face-mask while no interacting) are assumed to behave actively no matter what the others do, while N-types’ (those without face-mask while no interacting) do it only when they observe another node doing it, i.e., when at least an A-type’s is in the group. In this setup, any transmission may be avoided when interacting in groups of three, whereas all the N-type agents may be infected when interacting in pairs. Neglecting the group organization of contacts can thus lead to radically different outcomes, proving the need for a higher-order representation to account for context.
    }
    \label{fig_1}
\end{figure}

Figure \ref{fig_1} provides an illustrative example of the model, while showing the inadequacy of a pairwise representation for context-dependent processes and the consequent need for higher-order representations. Indeed, since the success of each pairwise transmission depends on the behaviors taken on by all the agents in the group, the exact description of the whole process requires knowledge of either the interaction structure (i.e., who gathers with whom, and when) and the functional dependence of the agents' behavior on their type and on the type-composition of a group. The exact specification of the interaction structure requires a huge amount of detailed information. While this is largely unavailable in most of the cases, we are anyway forced to resort to some approximate description if we want to gain clear knowledge from analytical results.

\section{MEAN-FIELD THEORY FOR MIXING IN GROUPS}
\label{sec:modeling}

The coarsest approximation is based on the well-mixed hypothesis, which, assuming that any agent can interact with any other one in the population, allows us to substitute local state variables with global aggregates. Nonetheless, we can enrich and extend this scheme by making the probability for a given interaction to occur be a function of the type of the agents it involves. The structure is thus described by means of probability mass functions providing the probability for each type to participate to groups of various size and type-composition. Put differently, we operate a dimensional reduction based on the assumption that the type of the agents is the sole property needed to infer the interaction structure. By taking into account the way the behavioral types mix, such description allows us to preserve the essential elements that determine the spreading process.

We first present the general mean-field theory valid for any group size, and then we provide explicit formulas for rank-$3$ hypergraphs.

\subsection{General case}
\label{sec:modeling_gen}

Consider a population (a set) of $N$ agents (nodes). Let $\rho$ be the proportion of A-type agents, hence $1-\rho$ the proportion of N-type's. We assume here that, at each time step, groups form at random, with probabilities depending solely on the types of the agents involved. Accordingly, the value of ${A_t}^{(n)}_{i_1,\dots,i_n}$ depends only on the set $\{\text{X}_{i_1},\dots,\text{X}_{i_n}\}$ of the agents' types, and we replace it with its expected value over the ensemble of hypergraphs parametrized by $N$, $\rho$, and a set of mixing parameters (defined below) regulating the average level of type-(dis)assortativity in a group. To this end, we need an expression for the number $g_{n_\text{A},n_\text{N}}^{(n)}$ of $\left(n_\text{A}+n_\text{N}\right)$-edges composed of $n_\text{A}$ A-type nodes and $n_\text{N}$ N-type ones, being subsets of $n$-edges (i.e., $n_\text{A}+n_\text{N}\leqslant n\leqslant n_\text{max}$), formed at each time step. From this we can then derive an expression for the probability that an agent of a certain type takes part in a group of any given type-composition.

Denoted with $e^{(n)}$ the number of $n$-edges formed at each time step, $g_{n_\text{A},n_\text{N}}^{(n)}$ is related to it through
\begin{equation}
p_{n_\text{A},n_\text{N}}^{(n)}=\frac{\ds g_{n_\text{A},n_\text{N}}^{(n)}}{\ds \binom{n}{n_\text{A}+n_\text{N}}e^{(n)}}\ ,
\label{eq:p}
\end{equation}

\noindent being $p_{n_\text{A},n_\text{N}}^{(n)}$ the probability that sampling $n_\text{A}+n_\text{N}$ nodes from a randomly chosen $n$-edge (there are $\binom{n}{n_\text{A}+n_\text{N}}e^{(n)}$ ways of doing it), the sample consists of $n_\text{A}$ A-type's and $n_\text{N}$ N-type's. Also, let $p_{l_\text{A},l_\text{N}\vert m_\text{A},m_\text{N}}^{(n)}$ be the conditional probability that, given a set of $m_\text{A}$ A-type and $m_\text{N}$ N-type nodes in an $n$-edge, sampling a set of other $l_\text{A}+l_\text{N}$ nodes in it ($m_\text{A}+l_\text{A}+m_\text{N}+l_\text{N}\leqslant n$), the sample consists of $l_\text{A}$ A-type's and $l_\text{N}$ N-type's.
Since there are $\binom{m_\text{X}+l_\text{X}}{m_\text{X}}$ ways of choosing $m_\text{X}$ nodes out of $m_\text{X}+l_\text{X}$, $\text{X}\in\{\text{A},\text{N}\}$, and $\binom{n - m_\text{A}-m_\text{N}}{l_\text{A}+l_\text{N}}$ ways of choosing $l_\text{A}+l_\text{N}$ nodes among the $n - m_\text{A}-m_\text{N}$ remaining ones, we get
\begin{align}
    p_{l_\text{A},l_\text{N}\vert m_\text{A},m_\text{N}}^{(n)} &= \frac{\displaystyle\binom{m_\text{A}+l_\text{A}}{m_\text{A}}\binom{m_\text{N}+l_\text{N}}{m_\text{N}}}{\displaystyle\binom{n - m_\text{A}-m_\text{N}}{l_\text{A}+l_\text{N}}}\, \frac{g_{m_\text{A}+l_\text{A},m_\text{N}+l_\text{N}}^{(n)}}{g_{m_\text{A},m_\text{N}}^{(n)}}\ .
    \label{eq:pn}
\end{align}

\noindent In particular, $p_{l_\text{A},l_\text{N}\vert 1,0}^{(n)}$ ($p_{l_\text{A},l_\text{N}\vert 0,1}^{(n)}$) is the sought probability that an A-type (N-type) node takes part in a group of size $n$ which includes other $l_\text{A}$ A-types' and $l_\text{N}$ N-types'. Therefore, indicated with $k_{1,0}^{(n)}$ ($k_{0,1}^{(n)}$) the expected $n$-degree of an A-type (N-type) node, i.e., the expected number of groups of size $n$ to which it takes part in per time step, and given $l_\text{A}+l_\text{N}=n-1$, then $\kappa_{l_\text{A},l_\text{N}\vert 1,0}^{(n)} = p_{l_\text{A},l_\text{N}\vert 1,0}^{(n)} k_{1,0}^{(n)}$ ($\kappa_{l_\text{A},l_\text{N}\vert 0,1}^{(n)} = p_{l_\text{A},l_\text{N}\vert 0,1}^{(n)} k_{0,1}^{(n)}$) is the expected number of groups of size $n$ in which an A-type (N-type) takes part with other $l_\text{A}$ A-type's and $l_\text{N}$ N-type's. These numbers, beside $N$ and $\rho$, specify completely the interaction structure.

To compute them, we need an expression for $g_{n_\text{A},n_\text{N}}^{(n)}$. For mixed subsets, i.e., with $n_\text{A},n_\text{N}\geqslant 1$, we have (see Appendix~\ref{sec:app_modeling_g} for the detailed derivation)
\begin{align}
    \notag g_{n_\text{A},n_\text{N}}^{(n)} =& \ \frac{N_{0,1} k_{0,1}^{(n)}}{n_\text{N}} \binom{n-1}{n_\text{A}+n_\text{N}-1} \binom{n_\text{A}+n_\text{N}-1}{n_\text{A}} \\
    \notag &\times p_{0,1\vert 0,1}^{(n)} p_{0,1\vert 0,2}^{(n)}\cdots p_{0,1\vert 0,n_\text{N}-1}^{(n)} \\
    &\times p_{1,0\vert 0,n_\text{N}}^{(n)} p_{1,0\vert 1,n_\text{N}}^{(n)} \cdots p_{1,0\vert n_\text{A}-1,n_\text{N}}^{(n)}\ ,
    \label{eq:g}
\end{align}

\noindent and for uniform N-type and A-type subsets,
\begin{align}
    g_{0,n_\text{N}}^{(n)} =& \ \frac{N_{0,1} k_{0,1}^{(n)}}{n_\text{N}} \binom{n-1}{n_\text{N}-1} \ p_{0,1\vert 0,1}^{(n)} p_{0,1\vert 0,2}^{(n)} \cdots p_{0,1\vert 0,n_\text{N}-1}^{(n)}\ ,
    \label{eq:g_N}
    \\
    g_{n_\text{A},0}^{(n)} =& \ \frac{N_{1,0} k_{1,0}^{(n)}}{n_\text{A}} \binom{n-1}{n_\text{A}-1} \ p_{1,0\vert 1,0}^{(n)} p_{1,0\vert 2,0}^{(n)} \cdots p_{1,0\vert n_\text{A}-1,0}^{(n)}\ .
    \label{eq:g_A}
\end{align}

Inserting Eqs.~(\ref{eq:g}) to~(\ref{eq:g_A}) into Eq.~(\ref{eq:pn}) we are thus expressing $p_{l_\text{A},l_\text{N}\vert m_\text{A},m_\text{N}}^{(n)}$ only in terms of conditional probabilities of finding a single node of given type, i.e., of the form $p_{1,0\vert m_\text{A},m_\text{N}}^{(n)}$ and $p_{0,1\vert m_\text{A},m_\text{N}}^{(n)}$. The latter are the only free variables left. To close the theory, given $m_\text{A}+m_\text{N}=m\leqslant n-1$, we parametrize them as follows:
\begin{align}
    p_{1,0\vert m_\text{A},m_\text{N}}^{(n)} &= \alpha_{m,m_\text{A}}^{(n)} \rho\ , & \text{if } m_\text{A} \leqslant m_\text{N}\ , \label{eq:pn1}
    \\
    p_{0,1\vert m_\text{A},m_\text{N}}^{(n)} &= \alpha_{m,m_\text{A}}^{(n)} \left(1-\rho\right)\ , & \text{if }  m_\text{A} > m_\text{N}\ . \label{eq:pn2}
\end{align}

\noindent Their counterparts are found from normalization, e.g., $p_{0,1\vert m_\text{A},m_\text{N}}^{(n)}=1-\alpha_{m,m_\text{A}}^{(n)} \rho$ if $m_\text{A} \leqslant m_\text{N}$. Parameter $\alpha_{m,m_\text{A}}^{(n)}$ governs the mixing probability in a $(m+1)$-edge, subset of an $n$-edge, conditioned on the presence of $m_\text{A}\leqslant m$ A-type nodes in it: $\alpha_{m,m_\text{A}}^{(n)}=1$ corresponds to homogeneous mixing, in which case it is only the proportion of the types in the population to determine the expected mixing; for $m_\text{A}\neq m_\text{N}$, $\alpha_{m,m_\text{A}}^{(n)}<1$ ($\alpha_{m,m_\text{A}}^{(n)}>1$) indicates (dis)assortativity of the majority in the subset towards the remaining node, which is therefore more (less) likely than expected to be of the same type of the majority; for $m_\text{A} = m_\text{N}$ (occurring only for $m$ even), $\alpha_{m,m_\text{A}}^{(n)}<1$ ($\alpha_{m,m_\text{A}}^{(n)}>1$) indicates asymmetric preference towards N-type (A-type) nodes.

Since $m_\text{A}\in\{0,\dots,m\}$, there are $m+1$ parameters characterizing the mixing within $(m+1)$-edges. However, only one of them is free, eventually implying that the interaction structure is uniquely determined by at most $\binom{n_\text{max}}{2}$ parameters (see Appendix~\ref{sec:app_modeling_params} for proof).

Part of our analysis will focus on how the dynamics is affected by how type-assortativity is distributed among the various group sizes. To this end, let us define the level of pairwise homophily (or just ``homophily'') in $n$-edges, $h^{(n)}$, as the probability that two agents are of the same type when interacting within a group of size $n$,
\begin{align}
    \notag h^{(n)} &= \frac{N_{1,0}k_{1,0}^{(n)}p_{1,0\vert 1,0}^{(n)} + N_{0,1}k_{0,1}^{(n)}p_{0,1\vert 0,1}^{(n)}}{N_{1,0}k_{1,0}^{(n)} + N_{0,1}k_{0,1}^{(n)}} \\
    &= 1 - \rho\left(1-\rho\right)\frac{\alpha_{1,0}^{(n)}k_{0,1}^{(n)}+\alpha_{1,1}^{(n)} k_{1,0}^{(n)}}{k^{(n)}} \,
    \label{eq:h_n}
\end{align}

\noindent where $k^{(n)} = \rho k_{1,0}^{(n)} + \left(1-\rho\right)k_{0,1}^{(n)}$ is the average $n$-degree. Note that this is only one of the several ways in which assortativity can be quantified within groups of size $n>2$. Indeed, one can consider a weaker notion of assortativity by defining an index for each one of the group compositions where one type is majoritarian (e.g., for triads, in addition to the index counting configurations where the three nodes are of the same type, another one counting those where two of them are of the same type)~\cite{veldt2021higher}. In this sense, assortativity in pairs is exceptional, for there is only one way of being majoritarian.
We choose to focus on pairwise assortativity because the direct interactions we will consider ---those mediating the spreading--- are pairwise, hence knowing how frequently different pairs form is of primary importance here. Nonetheless, from the previous analysis we know that fixing $h^{(n)}$ alone is not sufficient to fully specify the group organization. Therefore, unless otherwise specified, we fix the higher-order mixing parameters ($\alpha_{m,m_\text{A}}^{(n)}$, $m>1$) and vary instead the pairwise ones ($\alpha_{1,m_\text{A}}^{(n)}$, $m_\text{A}=0,1$). In particular, chosen a value for the average homophily $h=\sum_{n=2}^{n_{\text{max}}}h^{(n)} \left(n-1\right)k^{(n)}/k$, being $k=\sum_{n=2}^{n_{\text{max}}}\left(n-1\right)k^{(n)}$ the average pairwise degree (i.e., the average degree over the graph projection of the hypergraph), we vary the type-assortativities $\left\{h^{(n)}\right\}$ and analyze how this affects the dynamics. Using Eq.~(\ref{eq:h_n}), we find
\begin{equation}
    h=1 - (\alpha_{1,0}+\alpha_{1,1})\rho\left(1-\rho\right)\ ,
\end{equation}
where $\alpha_{1,0}=\sum_{n=2}^{n_{\text{max}}}\alpha_{1,0}^{(n)}\left(n-1\right)k_{0,1}^{(n)}/k = \sum_{n=2}^{n_{\text{max}}}\alpha_{1,1}^{(n)}\left(n-1\right)k_{1,0}^{(n)}/k=\alpha_{1,1}$ is the average pairwise mixing parameter.

\subsection{Pairs and triads}
\label{sec:modeling_3}

As a minimal application of this formalism, we will consider agents interacting only within pairs ($2$-edges) and triads ($3$-edges). To start with, Eqs.~(\ref{eq:pn1}) and (\ref{eq:pn2}) read
\begin{align}
    p_{1,0\vert 0,1}^{(2)} &= \alpha_{1,0}^{(2)}\rho\ ,
    \label{eq:p2_1} \\
    p_{0,1\vert 1,0}^{(2)} &= \alpha_{1,1}^{(2)}\left(1-\rho\right)\ ,
    \label{eq:p2_2}
\end{align}

\noindent for $n=2$, and
\begin{align}
    p_{1,0\vert 0,1}^{(3)} &= \alpha_{1,0}^{(3)}\rho\ ,
    \label{eq:p3_1} \\
    p_{0,1\vert 1,0}^{(3)} &= \alpha_{1,1}^{(3)}\left(1-\rho\right)\ ,
    \label{eq:p3_2} \\
    p_{1,0\vert 0,2}^{(3)} &= \alpha_{2,0}^{(3)}\rho\ ,
    \label{eq:p3_3} \\
    p_{1,0\vert 1,1}^{(3)} &= \alpha_{2,1}^{(3)}\rho\ ,
    \label{eq:p3_4} \\
    p_{0,1\vert 2,0}^{(3)} &= \alpha_{2,2}^{(3)}\left(1-\rho\right)\ ,
    \label{eq:p3_5}
\end{align}

\noindent for $n=3$. Choosing then $\alpha_{1,0}^{(2)}$, $\alpha_{1,0}^{(3)}$ and $\alpha_{2,1}^{(3)}$ as the free parameters, one finds the relations
\begin{align}
    \alpha_{1,1}^{(n)} &=  \frac{k_{0,1}^{(n)}}{k_{1,0}^{(n)}}\alpha_{1,0}^{(n)} \ \ (n = 2,3)\ ,\label{eq:alpha2_1,1}\\
    \alpha_{2,0}^{(3)} &=  \frac{\alpha_{1,0}^{(3)}\left(1-\alpha_{2,1}^{(3)}\rho\right)}{1-\alpha_{1,0}^{(3)}\rho}\ ,\label{eq:alpha3_2,0}\\
    \alpha_{2,2}^{(3)} &= \frac{\alpha_{1,1}^{(3)}\alpha_{2,1}^{(3)}\rho}{1-\alpha_{1,1}^{(3)}\left(1-\rho\right)}\ . \label{eq:alpha3_3,2}
\end{align}

\noindent We eventually find the following expressions for the mixing probabilities in triads conditioned on a single node,
\begin{align}
    p_{2,0\vert 0,1}^{(3)} &= \alpha_{1,0}^{(3)}\rho ~\alpha_{2,1}^{(3)}\rho\ , \label{eq:p'3_1}\\
    p_{1,1\vert 0,1}^{(3)} &= 2\alpha_{1,0}^{(3)}\rho\left(1-\alpha_{2,1}^{(3)}\rho\right)\ , \label{eq:p'3_2}\\
    p_{0,2\vert 0,1}^{(3)} &= \left(1-\alpha_{1,0}^{(3)}\rho\right)\left(1-\alpha_{2,0}^{(3)}\rho\right)\ , \label{eq:p'3_3}\\
    p_{0,2\vert 1,0}^{(3)} &= \alpha_{1,1}^{(3)}\left(1-\rho\right)\left(1-\alpha_{2,1}^{(3)}\rho\right)\ , \label{eq:p'3_4}\\
    p_{1,1\vert 1,0}^{(3)} &= 2\alpha_{1,1}^{(3)}\left(1-\rho\right)\alpha_{2,1}^{(3)}\rho\ , \label{eq:p'3_5}\\
    p_{2,0\vert 1,0}^{(3)} &= \left[1-\alpha_{1,1}^{(3)}\left(1-\rho\right)\right]\left[1-\alpha_{2,2}^{(3)}\left(1-\rho\right)\right]\ , \label{eq:p'3_6}
\end{align}

\noindent which, together with Eqs.~(\ref{eq:p2_1}) to~(\ref{eq:p3_2}), specify completely the mixing patterns for a single node within groups of size $n=2,3$. Note that, to ensure that all the mixing probabilities lie in $\left[0,1\right]$, the intervals of variation of $\alpha_{1,0}^{(2)}$, $\alpha_{1,0}^{(3)}$ and $\alpha_{2,1}^{(3)}$ must fulfill the following constraints:
{\small
\begin{align}
    \alpha_{1,0}^{(2)}&\in\left[0,\text{min}\left\{\frac1\rho,\frac1{1-\rho},\frac{k_{1,0}^{(2)}}{\left(1-\rho\right)k_{0,1}^{(2)}}\right\}\right]\ , \label{eq:lim1} \\
    \alpha_{1,0}^{(3)}&\in\left[0,\text{min}\left\{\frac1\rho,\frac1{1-\rho},\frac{k_{1,0}^{(3)}}{\left(1-\rho\right)k_{0,1}^{(3)}},\frac{k^{(3)}}{3\rho\left(1-\rho\right)k_{0,1}^{(3)}}\right\}\right]\ , \label{eq:lim2} \\
    \alpha_{2,1}^{(3)}&\in\left[\text{max}\left\{0,\frac{2\alpha_{1,0}^{(3)}\rho-1}{\alpha_{1,0}^{(3)}\rho^2}\right\},\text{min}\left\{\frac1\rho,\frac{1-\alpha_{1,1}^{(3)}\left(1-\rho\right)}{\alpha_{1,1}^{(3)}\rho\left(1-\rho\right)}\right\}\right]\ . \label{eq:lim3}
\end{align}}

\noindent In particular, the forth upper bound in Eq.~(\ref{eq:lim2}) guarantees the feasibility of the structure by constraining the lower bound for $\alpha_{2,1}^{(3)}$ to stay below its upper bound.

When keeping fixed the average homophily $h$ (hence $\alpha_{1,0}$), we get the additional relation
\begin{equation}
    \alpha_{1,0}^{(3)} = \frac{1}{2k_{0,1}^{(3)}}\left(\alpha_{1,0}k - \alpha_{1,0}^{(2)}k_{0,1}^{(2)}\right)\ ,
    \label{eq:alpha3_1,0}
\end{equation}

\noindent with $k=k^{(2)}+2k^{(3)}$. Equation~(\ref{eq:alpha3_1,0}) puts further bounds on $\alpha_{1,0}^{(2)}$ in order for $\alpha_{1,0}^{(3)}$ to satisfy Eq.~(\ref{eq:lim2}).

Dealing with pairs and triads only, it is convenient to switch to a more explicit notation from now on. In particular, we use $k_{\text{N}}^{(n)}$ instead of $k_{0,1}^{(n)}$, $n=2,3$; $\kappa_{\text{N}\vert\text{N}}$ and $\kappa_{\text{A}\vert\text{N}}$ instead of $\kappa_{0,1\vert 0,1}^{(2)}$ and $\kappa_{1,0\vert 0,1}^{(2)}$, respectively; and $\kappa_{\text{N,N}\vert\text{N}}$, $\kappa_{\text{A,N}\vert\text{N}}$ and $\kappa_{\text{A,A}\vert\text{N}}$ instead of $\kappa_{0,2\vert 0,1}^{(3)}$, $\kappa_{1,1\vert 0,1}^{(3)}$ and $\kappa_{2,0\vert 0,1}^{(3)}$, respectively (analogously when exchanging N with A). Also, we denote the three free parameters we have as $\alpha_2\equiv\alpha_{1,0}^{(2)}$, $\alpha_3\equiv\alpha_{1,0}^{(3)}$ and $\beta_{3,1}\equiv\alpha_{2,1}^{(3)}$, and the average pairwise mixing parameter as $\alpha \equiv \alpha_{1,0} = \alpha_{1,1}$.

\section{CONTEXT-DEPENDENT BEHAVIOR}
\label{sec:behavioral}

The next step is to model the behavioral difference between the two types.
The probability of actively modifying the spread can be thought of as consisting of a type-specific, context-independent part (e.g., prosociality~\cite{eisenberg2006prosocial}) and a context-dependent one, i.e., relying on the observed behavior of the others in the group~\cite{asch1955opinions}. The probability of being active is thus a dynamic object. In this regard, a simplifying (but, for some applications, seemingly realistic) assumption is that such probability converges to an equilibrium value during a timescale which is much shorter than the duration of a group interaction (e.g., the decision of wearing a face-mask may be considered to be made up in the very first moments of a gathering). With this timescale separation, the probability of being active is assumed to instantly attain its equilibrium value, effectively becoming a time-independent quantity.

Now, given a group of size $n$, let $q_{\text{X}}^{(n_\text{X}-1,n-1)}$ be the probability that a X-type agent adopts an active behavior when $n_\text{X}-1$ out of the other $n-1$ agents in the group are of X-type's as well. In accordance to the meaning we associated with the labels we require that (i) the probability of behaving actively, within a mixed group, is always higher for an A-type agent than for a N-type's; and (ii) such probability increases with the number of A-type's present in the group. Taken together they imply the following chain of inequalities,
\begin{equation}
    q_{\text{A}}^{(n-1,n-1)} \geqslant q_{\text{A}}^{(n_\text{A}-1,n-1)} \geqslant q_{\text{N}}^{(n_\text{N}-1,n-1)} \geqslant q_{\text{N}}^{(n-1,n-1)}\ ,
    \label{eq:req}
\end{equation}

\noindent for $1 \leqslant n_\text{A}\leqslant n-1$, $n_\text{N} = n - n_\text{A}$. Equation~(\ref{eq:req}) defines the qualitative behavioral difference between the types.

Having access to empirical measurements of the $q_\text{X}$s, one could simply fit them in the dynamic equations. In the absence of empirical information, we may suppose different functional forms for them, satisfying Eq.~(\ref{eq:req}). One possibility is to let the $q_\text{X}$s emerge dynamically as a result of a behavioral adaptation, as shown in Section~\ref{sec:social contagion}. Another one, presented in Section~\ref{sec:binary}, is to postulate binary functional forms for the $q_\text{X}$s, where agents of each type are active or inactive in a deterministic way depending on the composition of the groups they take part in. The simplicity of the binary forms will allow us to obtain explicit analytical results, gaining essential insights to interpret more complicated scenarios.

To notice that any of the behavioral dynamics proposed here can be easily generalized to accommodate additional potential factors, such as external sources of influence (e.g., mass media), seen as a background context modifying the baseline probability of adoption, or some coupling with the spreading dynamics (e.g., awareness raised by information on an epidemics).

\subsection{Social contagion model}
\label{sec:social contagion}

Here we let agents' behavior emerge through a process of social contagion~\cite{hill2010infectious,Centola2010spread,hebert2022source} modelled as a susceptible-infectious-susceptible (SIS) dynamics, where here \lq S\rq\ and \lq I\rq\ represent nonadoption and adoption of the active behavior, respectively, with additional endogenous transitions moving agents from state S to state I and vice versa at rates depending solely on their type. These rates quantify some cost of adopting an active behavior (e.g., the discomfort of wearing a face-mask) and the will to bear that cost (e.g., due to prosociality or vulnerability to a disease). Denoting with $c_\text{X}$ the I-to-S rate and with $b_\text{X}$ the S-to-I rate, $\text{X}\in\left\{\text{A},\text{N}\right\}$, we describe the behavioral dynamics via the following system (see the derivation in Appendix~\ref{sec:behav_dyn})
\begin{align}
    \dot{q_{\text{A}}} &= \frac{n_\text{N}}{n-1}\left[q_{\text{N}}-q_{\text{A}}\right] + b_{\text{A}}\left[1-q_{\text{A}}\right] - c_{\text{A}}q_{\text{A}}\ , \label{eq:qA} \\
    \dot{q_{\text{N}}} &= \frac{n_\text{A}}{n-1}\left[q_{\text{A}}-q_{\text{N}}\right]
    + b_{\text{N}}\left[1-q_{\text{N}}\right] - c_{\text{N}}q_{\text{N}}\ . \label{eq:qN}
\end{align}

\noindent where we denoted $q_{\text{X}}^{(n_\text{X}-1,n-1)}(t)$ simply as $q_{\text{X}}$ for a lighter notation. The first term of each of the two equations contains the context-dependent dynamics, which has the form of a linear consensus formation process~\cite{degroot1974reaching}; the remaining terms contain the context-independent, type-specific dynamics.

The system admits a unique solution for each possible composition of a group (see Appendix~\ref{sec:behav_dyn} for details). Interestingly, for a wide range of values of the transition rates ($b$s and $c$s) producing a significant behavioral difference between the types ---which is what really motivates their definition---, the equilibrium probability of adoption shows a marked nonlinear dependence on the number of A-type (or N-type) agents in the group. This is especially true for $q_\text{N}$ ($q_\text{A}$ is always high), growing nonlinearly from low values in a uniform N-type group ($n_\text{A}=0$) to medium-high values when all the others in the group are A-type's ($n_\text{A}=n-1$).

\subsection{Binary models}
\label{sec:binary}

We define here the binary models we call of \lq easy adaptation\rq\ and \lq hard adaptation\rq. In both of them $q_\text{A}=1$ always, i.e., A-type agents are assumed to behave according to their intrinsic propensity and independently of the context observed in a group. We then have the following: \\
\indent (i) In easy adaptation, $q_\text{N}$ is $1$ when at least an A-type is present (specifically, for rank-$3$ hypergraphs, in \{A,N\}, \{A,N,N\} and \{A,A,N\} groups)) and $0$ otherwise (that is, in uniform N-type groups). N-type agents are easily (maximally) driven by A-type's in this scenario; \\
\indent (ii) In hard adaptation, $q_\text{N}$ is $1$ when all the other individuals in the group are A-type's (\{A,N\} and \{A,A,N\} groups) and is $0$ otherwise., i.e., N-type's are hardly (minimally) driven by A-type's in this case.

The only and crucial difference between the two behavioral models is, for rank-$3$ hypergraphs, the N-type agents' behavior in \{A,N,N\} groups: the presence of the A-type agent induces active behavior in the two N-type's in the easy adaptation model, but not in the hard adaptation one.
As shown in the next sections, this difference modifies the spreading process by making it strongly dependent on the behavioral dynamics taking place at group-level, hence on the properties of the higher-order interaction structure.

\section{SPREADING DYNAMICS}
\label{sec:spreading}

Once the interaction structure and the behavior's context-dependency have been modelled, we are ready to investigate how they jointly affect a spreading process taking place upon that structure. For the sake of simplicity, we assume the spreading process to be represented by a SIS dynamics. The latter is a basic model in the study of epidemics, which is precisely the specific application for which the mathematical machinery we developed will be used below. Without loss of generality, we will therefore make use of an epidemiological terminology.

Let thus $\lambda$ be the transmission rate and $\mu$ the recovery rate. We denote with $Y_{\text{X}} \equiv Y_{\text{X}}(t)$ the fraction of agents of type $\text{X}\in\left\{\text{A},\text{N}\right\}$ in compartment $Y\in\left\{S,I\right\}$ at time $t$. In particular, $I_{\text{X}}$ is the type-specific prevalence for type X, while $I = \rho I_{\text{A}} + (1-\rho) I_{\text{N}}$ is the prevalence overall. The dynamics can be modelled through the following system of differential equations
\begin{align}
    \dot{I_{\text{A}}} =& -\mu I_{\text{A}} + \lambda S_{\text{A}}\left(I_{\text{A}}\theta_{\text{A}\rightarrow\text{A}} + I_{\text{N}}\theta_{\text{N}\rightarrow\text{A}}\right)\ , \label{eq:dynA}\\
    \dot{I_{\text{N}}} =& -\mu I_{\text{N}} + \lambda S_{\text{N}}\left(I_{\text{A}}\theta_{\text{A}\rightarrow\text{N}} + I_{\text{N}}\theta_{\text{N}\rightarrow\text{N}}\right)\ ,  \label{eq:dynN}
\end{align}

\noindent with $S_{\text{X}} = 1- I_{\text{X}}$. Here $\lambda\theta_{\text{X}\rightarrow\text{Z}}$ is the total transmission rate from a X-type to a $\text{Z}$-type, reading
\begin{align}
    \notag \theta_{\text{A}\rightarrow\text{A}} =&\ \kappa_{\text{A}\vert \text{A}} r_{\text{A}}^{(1,1)}s_{\text{A}}^{(1,1)} \\ &+ \kappa_{\text{A,N}\vert \text{A}} r_{\text{A}}^{(1,2)}s_{\text{A}}^{(1,2)} + 2\kappa_{\text{A,A}\vert \text{A}} r_{\text{A}}^{(2,2)}s_{\text{A}}^{(2,2)}\ , \label{eq:thetaAA} \\
    \notag \theta_{\text{N}\rightarrow\text{A}} =&\ \kappa_{\text{N}\vert \text{A}} r_{\text{N}}^{(0,1)}s_{\text{A}}^{(0,1)} \\ &+ \kappa_{\text{A,N}\vert \text{A}} r_{\text{N}}^{(0,2)}s_{\text{A}}^{(1,2)} + 2\kappa_{\text{N,N}\vert \text{A}} r_{\text{N}}^{(1,2)}s_{\text{A}}^{(0,2)}\ , \label{eq:thetaNA} \\
    \notag \theta_{\text{A}\rightarrow\text{N}} =&\ \kappa_{\text{A}\vert \text{N}} r_{\text{A}}^{(0,1)}s_{\text{N}}^{(0,1)} \\ &+ \kappa_{\text{A,N}\vert \text{N}} r_{\text{A}}^{(0,2)}s_{\text{N}}^{(1,2)} + 2\kappa_{\text{A,A}\vert \text{N}} r_{\text{A}}^{(1,2)}s_{\text{N}}^{(0,2)}\ , \label{eq:thetaAN} \\
    \notag \theta_{\text{N}\rightarrow\text{N}} =&\ \kappa_{\text{N}\vert \text{N}} r_{\text{N}}^{(1,1)}s_{\text{N}}^{(1,1)} \\ &+ \kappa_{\text{A,N}\vert \text{N}} r_{\text{N}}^{(1,2)}s_{\text{N}}^{(1,2)} + 2\kappa_{\text{N,N}\vert \text{N}} r_{\text{N}}^{(2,2)}s_{\text{N}}^{(2,2)}\ , \label{eq:thetaNN}
\end{align}

\noindent where $\lambda r_{\text{X}}^{(n_\text{X}-1,n-1)}s_{\text{Z}}^{(n_\text{Z}-1,n-1)}$ is the effective transmission rate for an interaction occurring within a $n$-edge composed of $n_\text{X}$ X-type and $n_\text{Z}$ Z-type agents. This is written as the product of an out-going (i.e., depending on the infected, a X-type) transmission probability $r_{\text{X}}^{(n_\text{X}-1,n-1)}$ and an in-going (i.e., depending on the susceptible, a Z-type) transmission probability $s_{\text{Z}}^{(n_\text{Z}-1,n-1)}$. With the same notation used in Sec.~\ref{sec:behavioral}, the superscript $(n_\text{X}-1,n-1)$ indicates how many of the other $n-1$ nodes in the group are of type X (subscript), given at least an X-type is present. In other words, $r_\text{X}$ and $s_\text{X}$ encode the context within which a direct interaction occurs from the perspective of a X-type agent. Their form is dictated by the specific mechanism through which transmission can be modified and, via their dependence on the agents' probability to adopt active behavior, by how the mechanism is affected by the size and the type-composition of a group. For prophylactic mechanisms, $r_\text{X}$ and $s_\text{X}$ act as reduction factors of the transmission probability.

Going back to Eqs.~(\ref{eq:dynA}) and~(\ref{eq:dynN}) and linearizing them around the epidemic-free equilibrium, $(I_{\text{A}}, I_{\text{N}}, S_{\text{A}}, S_{\text{N}}) \approx (0,0,1,1)$, the associated Jacobian matrix reads
\begin{equation}
    \mathbf{J} = \begin{pmatrix}
    \lambda \theta_{\text{A}\rightarrow\text{A}} - \mu & \lambda \theta_{\text{N}\rightarrow\text{A}} \\
    \lambda \theta_{\text{A}\rightarrow\text{N}} & \lambda \theta_{\text{N}\rightarrow\text{N}} - \mu
    \end{pmatrix}\ ,
\end{equation}

\noindent i.e., $\mathbf{J} = \lambda \mathbf{\Theta} - \mu \mathbb{I}_{2\times2}$, being $\mathbf{\Theta}$ the matrix with entries $\left\{\theta_{\text{X}\rightarrow\text{Z}}\right\}$, and $\mathbb{I}_{2\times2}$ the $2\times2$ identity matrix. The basic reproduction number, $R$, is then given by the largest eigenvalue of the next generation matrix (NGM) \cite{diekmann2010construction}, $\text{NGM}=\left(\lambda/\mu\right)\mathbf{\Theta}$, as
\begin{equation}
    R = \frac{\lambda k_{\text{eff}}}{\mu}\ ,
\end{equation}

\noindent with

\begin{align}
    \notag k_{\text{eff}} &= \frac{1}{2}\bigg[ \theta_{\text{A}\rightarrow\text{A}}+\theta_{\text{N}\rightarrow\text{N}} \\
    &~~~~~~~+\sqrt{\left(\theta_{\text{A}\rightarrow\text{A}}-\theta_{\text{N}\rightarrow\text{N}}\right)^2 + 4\theta_{\text{N}\rightarrow\text{A}}\theta_{\text{A}\rightarrow\text{N}}}\,\bigg]\ ,
    \label{eq:k_eff}
\end{align}

\noindent being the largest eigenvalue of $\mathbf{\Theta}$, representing the effective ---for spreading--- average pairwise degree. Imposing $R=1$ provides the epidemic threshold $\lambda_\text{c} = \mu/k_{\text{eff}}$, above which the epidemic-free equilibrium is unstable and an endemic state is reached. If there is no modification of the transmission (i.e., $r_\text{X}=s_\text{X}=1$, $\forall \text{X}$) and there is no correlation between type and degree ($k_{0,1}^{(n)}=k_{1,0}^{(n)}=k^{(n)}$, $\forall n$), all agents are equivalent and mixing is irrelevant, hence $k_{\text{eff}} = k$, $R=R_0\equiv\lambda k/\mu$ ($\lambda_\text{c} = \mu/k$), and the homogeneous mean-field threshold of the SIS model on networks is recovered.

Notice that it is possible to get a closed form for the endemic solution of Eqs.~(\ref{eq:dynA}) and~(\ref{eq:dynN}). However, the complicated form of the solution does not allow to draw any noteworthy conclusion directly from it, hence we do not report its expression here. Nonetheless, an approximate solution that preserves the qualitative behavior of the system can be found under high prophylactic efficacy (see Appendix~\ref{sec:prev_approx}). Moreover, we anticipate that no qualitative changes are observed in the results presented below when, instead of a SIS dynamics, a susceptible-infectious-recovered (SIR) one is considered. In that case, the results refer to the final attack rate (total fraction of agents that got infected during an outbreak) rather than to the equilibrium prevalence.

\subsection{Application to face-masks adoption}

To describe an epidemic spreading in the presence of face-masks, we need to specify a suitable form for the probabilities $r_\text{X}$ and $s_\text{X}$ appearing in Eqs.~(\ref{eq:thetaAA}) to~(\ref{eq:thetaNN}). To this end, let $\varepsilon_{\text{out}}$, $\varepsilon_{\text{in}}\in[0,1]$ be the out- and in-going efficacy of a face-mask, respectively.
The quantities $\varepsilon_{\text{out}} q_{\text{X}}^{(n_\text{X}-1,n-1)}$ and $\varepsilon_{\text{in}} q_{\text{X}}^{(n_\text{X}-1,n-1)}$ are then the probabilities that a X-type individual wears a mask and that the mask avoids, respectively, out-going transmission when the X-type is infectious, and in-going transmission when the X-type is exposed to an infection. In more general terms, they quantify how strongly the spreading process can be affected by an agent's behavior: for perfect efficacy, the success of a spreading event is decided in a deterministic way by the eventually adopted behavior; on the contrary, for null efficacy, spreading and behavior fully decouple. Thus we have
\begin{align}
    r_{\text{X}}^{(n_\text{X}-1,n-1)} &= 1 - \varepsilon_{\text{out}} q_{\text{X}}^{(n_\text{X}-1,n-1)}\ , \label{eq:r_out}\\
    s_{\text{X}}^{(n_\text{X}-1,n-1)} &= 1 - \varepsilon_{\text{in}} q_{\text{X}}^{(n_\text{X}-1,n-1)}\ . \label{eq:r_in}
\end{align}

\noindent In the following we assume $\varepsilon_{\text{out}}=0.9$ and $\varepsilon_{\text{in}}=0.5$. The effect of varying each efficacy will be discussed whenever significant.

\begin{figure*}
    \centering
    \includegraphics[width = 1.\linewidth]{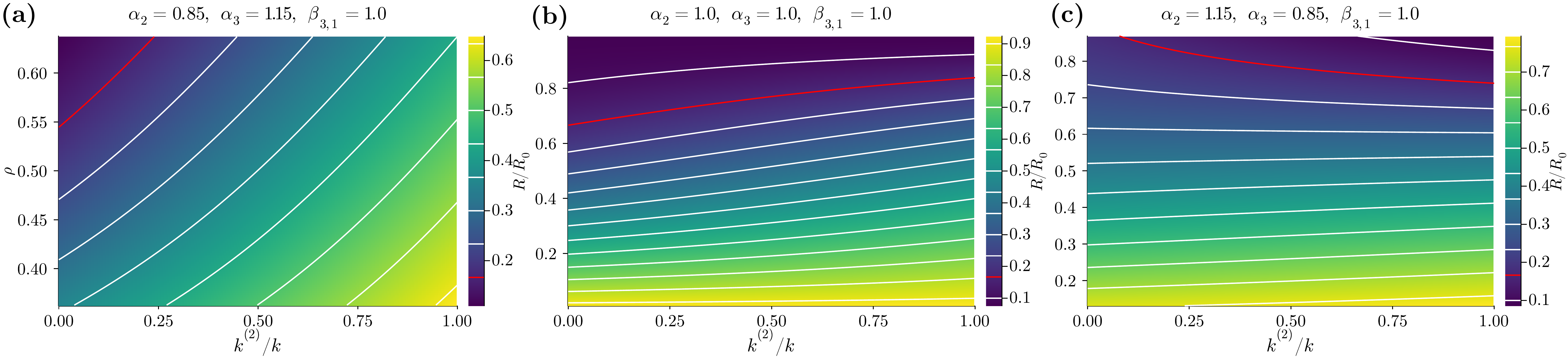}
    \caption{Normalized reproduction number, $R/R_0=k_\text{eff}/k$, computed from Eq.~(\ref{eq:k_eff}), as a function of the proportion of $2$-edges, $k^{(2)}/k$, and of the A-type proportion in the population, $\rho$, for $\beta_{3,1}=1.0$, $R_0 = k = 6$, and {\bf (a)} $\alpha_2 = 0.85$ and $\alpha_3 = 1.15$, {\bf (b)} $\alpha_2 = \alpha_3 = 1.0$, {\bf (c)} $\alpha_2 = 1.15$ and $\alpha_3 = 0.85$. Since $k=k^{(2)}+2k^{(3)}$, $k^{(2)}/k=0$ means nodes interacts only within triads, while $k^{(2)}/k=1$ means they do it just in pairs. White curves report the levels indicated in the respective colorbar. The red one indicates the critical curve $R=1$ ($R/R_0=1/6$), above which the disease-free state is stable.}
    \label{fig_2}
\end{figure*}

\section{RESULTS}
\label{sec:results}

To study how the group structure of the contacts affects the spreading dynamics, we vary the interaction structure along two directions. For given average values of degree, $k$, and assortativity, $\alpha$, we ask what is the effect of distributing (i) the degree ($k_\text{A}^{(n)}$ and $k_\text{N}^{(n)}$) and (ii) the assortativity ($\alpha_{1,0}^{(n)}$ and $\alpha_{1,1}^{(n)}$) differently at the varying of the group size $n$. To isolate the role of the mixing, we assume that type and degree are hereafter independent random variables. Therefore, a priori, $k_\text{A}^{(n)} = k_\text{N}^{(n)} = k^{(n)}$, $\forall n$, and consequently $\alpha_{1,1}^{(n)}=\alpha_{1,0}^{(n)}\equiv \alpha_n$. We consider individuals gathering in pairs and triads, i.e., $n=2,3$.

In the following, we display and analyze the results for the social contagion model of Sec.~\ref{sec:social contagion}. To this end, we leverage the basic understanding provided by the binary models of Sec.~\ref{sec:binary}. We fix the adoption probabilities by taking $c_{\text{A}}=c_{\text{N}}=0.05$, $b_{\text{A}} = 20 c_{\text{A}} = 1.0$ and $b_{\text{N}} = c_{\text{N}}/20 = 0.0025$, yielding a probability of adoption of around $95\%$ and $5\%$ in an uniform A-type and N-type group, respectively. In a $\{\text{A},\text{N}\}$ $2$-edge, the adoption probability jumps to around $87\%$ for the N-type agent and decreases to around $91\%$ for the A-type's. In a $\{\text{A},\text{N},\text{N}\}$ $3$-edge, it goes to around $80\%$ for the N-type's and to $88\%$ for the A-type's, while in a $\{\text{A},\text{A},\text{N}\}$ one, to around $89\%$ and $93\%$, respectively. Apart from the numerical details, what ultimately matters is that the A-type agents act as indirect modifiers of the infectious interactions in a group by lowering the transmission probability through the prophylaxis induced in the N-types'.

\subsection{Varying the degree distribution among group sizes}
\label{sec:results_contact}

Suppose that the mixing parameters for each group size, $\alpha_2$, $\alpha_3$ and $\beta_{3,1}$, are given. We then vary $k^{(2)}$ and $k^{(3)}$ respecting the constraint $k = k^{(2)}+2k^{(3)}$ and study how this affects the dynamics.

\subsubsection{Reproduction number}
\label{sec:results_contact_R}

\begin{figure*}
    \centering
    \includegraphics[width = 1.\linewidth]{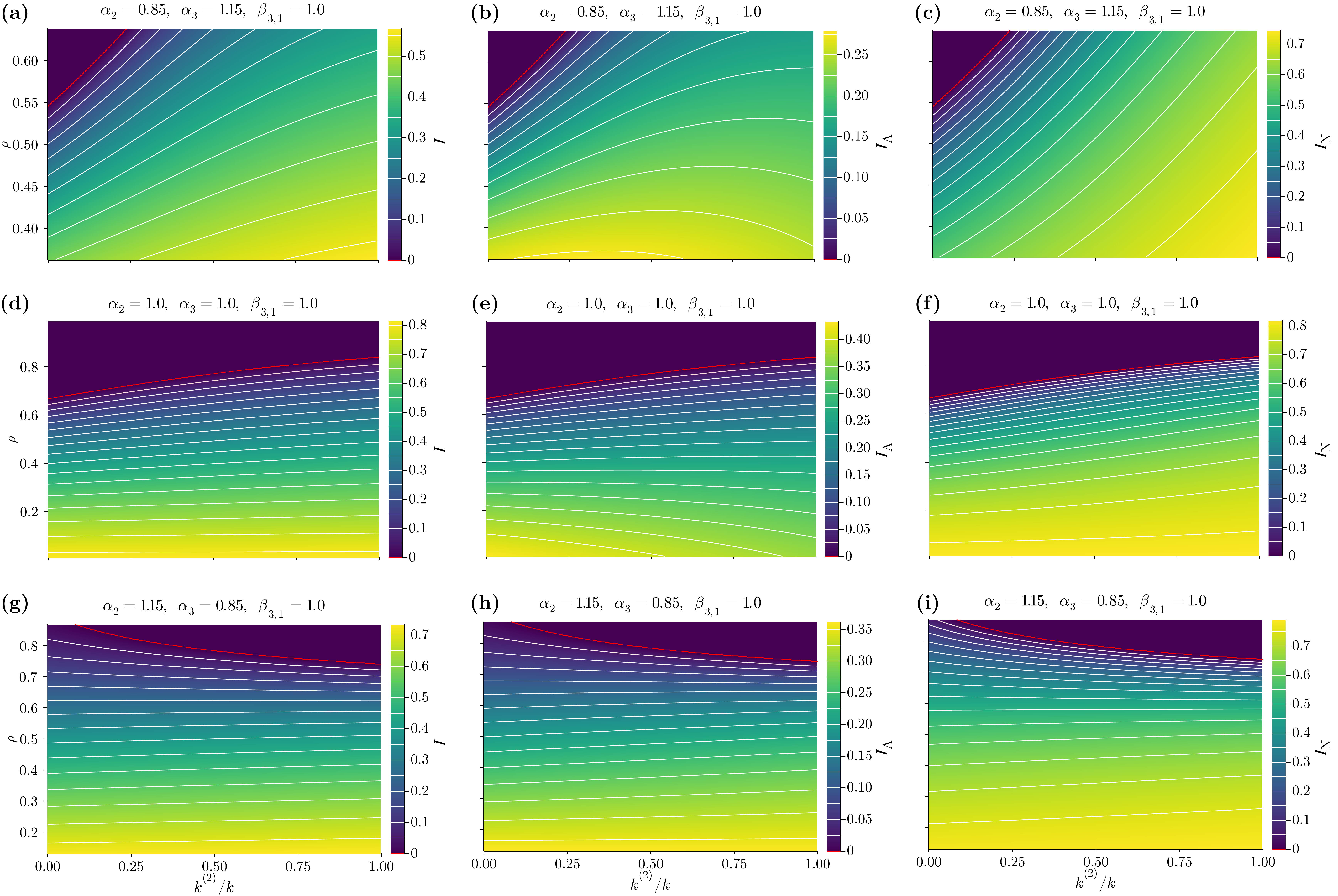}
    \caption{Equilibrium endemic state (fixed point of the system of Eqs.~(\ref{eq:dynA}) and~(\ref{eq:dynN})) as a function of the proportion of $2$-edges, $k^{(2)}/k$, and of the A-type proportion in the population, $\rho$, for $\beta_{3,1}=1.0$, $R_0 = k = 6$, and {\bf (a--c)} $\alpha_2 = 0.85$ and $\alpha_3 = 1.15$, {\bf (d--f)} $\alpha_2 = \alpha_3 = 1.0$, {\bf (g--i)} $\alpha_2 = 1.15$ and $\alpha_3 = 0.85$. {\bf (a, d, g)} Overall prevalence, $I$, {\bf (b, e, h)} prevalence for type A, $I_\text{A}$, and {\bf (c, f, i)} prevalence for type N, $I_\text{N}$. Since $k=k^{(2)}+2k^{(3)}$, $k^{(2)}/k=0$ means nodes interacts only within triads, while $k^{(2)}/k=1$ means they do it just in pairs. White curves report the levels indicated in the respective colorbar. The red one indicates the critical curve $R=1$, above which the disease-free state is stable.}
    \label{fig_3}
\end{figure*}

Let us first consider the basic reproduction number, $R$. The behavior of $R$ generally depends on both the mixing pattern ($\alpha_2$, $\alpha_3$, $\beta_{3,1}$) and the type-composition of the population ($\rho$), but in a different way depending on the system being closer to the easy or the hard adaptation scenario. This is observed, for example, for the intermediate scenario we used (see Fig.~\ref{fig_2}) ---of which the easy adaptation limit is a good proxy. Assuming an ideal perfect protection in at least one direction (i.e., $\varepsilon_\text{out}=1$ and/or $\varepsilon_\text{in}=1$), we provide explicit conditions for those dependencies (see Appendix~\ref{sec:R_contact} for derivation). In particular, in easy adaptation, there exists a threshold value
\begin{equation}
    \tilde\rho=\frac{2 \alpha_3 - \alpha_2}{\alpha_3\beta_{3,1}}\ ,
    \label{eq:rho_th_}
\end{equation}

\noindent at which a N-type individual has the same probability of finding at least an A-type's ---and so be induced to adoption--- in a pair or in a triad.
That probability becomes lower in a pair when $\rho<\tilde\rho$, leading $R$ to increase with $k^{(2)}$ (for fixed $k$); vice versa, it decreases with $k^{(2)}$ when $\rho>\tilde\rho$. As a consequence, expressed the critical threshold as the value $\rho=\rho_\text{c}$ at which $R=1$ (e.g., red curves in Fig.~\ref{fig_2}), the threshold moves up or down with $k^{(2)}$ depending on whether $\rho_\text{c} < \tilde\rho$ or $\rho_\text{c} > \tilde\rho$.
Notice that $\tilde\rho\in\left[0,1\right]$ requires $\alpha_3\in\left[\alpha_2/2,\alpha_2/\left(2-\beta_{3,1}\right)\right]$. When $\alpha_3<\alpha_2/2$, $\tilde\rho< 0$ and $R$ thus decreases with $k^{(2)}$, for it is easier for a N-type's to find at least an A-type's in a pair than in a triad, whatever the fraction $\rho$ of A-type's in the population. Said differently, the induced prophylaxis is less frequent in triads than in pairs because the former are comparatively too much assortative, therefore $R$ is minimized by interacting in pairs only. In the opposite case in which $\alpha_3>\alpha_2/\left(2-\beta_{3,1}\right)$, $\tilde\rho> 1$ and $R$ thus always increases with $k^{(2)}$, for a N-type's is always more likely to find at least an A-type's in a triad than in a pair in this case. Among the groups containing type A, while a \{A,A,N\} triad is dynamically equivalent to a \{A,N\} pair, a \{A,N,N\} triad has the advantage of weakening the N-N contact, with the result that $R$ is minimised by interacting in triads only.

Going back to the scenario considered in Fig.~\ref{fig_2}, we see that, treating it as if was an easy adaptation scenario with perfect unilateral protection, $\tilde\rho\geqslant1$ for $\alpha_2\leqslant\alpha_3$ (since $\beta_{3,1}=1$), while $\tilde\rho\approx 0.65$ for $\alpha_2 = 1.15$ and $\alpha_3 = 0.85$, which is anyway not far from the true threshold ($\rho\approx 0.60$). This is consistent with the results reported in Fig.~\ref{fig_2}, confirming the explanatory power of the simplified dynamics assumed to derive Eq.~(\ref{eq:rho_th_}).

In the hard adaptation limit, under the same assumptions (see Appendix~\ref{sec:R_contact}), the shape of $R$ is instead always monotonic, increasing or decreasing with $k^{(2)}$ solely depending on whether $\alpha_2$ is smaller or larger than $\alpha_3$, respectively. Indeed, since there is no indirect mechanism at work in the \{A,N,N\} triads in this case, reducing $R$ simply amounts to rise disassortativity and in turn increase the frequency of adoption for the N-type's.

All in all, this analysis proves that, depending on the behavioral properties of the agents, solely changing the group-size distribution can move the system from the subcritical (disease-free state) to the supercritical phase (endemic states), and vice versa.

\subsubsection{Prevalence}
\label{sec:results_contact_prev}

Looking now at the levels of prevalence at equilibrium, we can appreciate how differently the two types are affected by the interaction structure. We can first observe that $I_\text{N}$ qualitatively behaves like $R$ (compare for example Figs.~\ref{fig_3}(c), \ref{fig_3}(f) and \ref{fig_3}(i) with Fig.~\ref{fig_2}), therefore the analysis following Eq.~(\ref{eq:rho_th_}) approximately applies to $I_\text{N}$ too. This correspondence between $I_\text{N}$ and $R$ is actually expected whenever there is a substantial behavioral difference between the two types and prophylactic efficacy is high. In such case (see Appendix~\ref{sec:prev_approx} for proof) the epidemic pressure mainly comes from type N and the dynamics for the latter is well approximated by a one-type population SIS where both $I_\text{N}$ and $R$ are largely determined by $\theta_{\text{N}\rightarrow\text{N}}$. On the other hand, since $\theta_{\text{N}\rightarrow\text{A}}$ can show a different and opposite dependence on the parameters compared with $\theta_{\text{N}\rightarrow\text{N}}$, $I_\text{A}$ might not generally follow $R$, as seen, for instance, comparing Figs.~\ref{fig_3}(b), \ref{fig_3}(e) and \ref{fig_3}(h) with Fig.~\ref{fig_2}. The results for $I_\text{A}$ thus require a more detailed inspection. For the considered scenario, they broadly overlap with those obtained in the easy adaptation one (compare with Fig.~S2 of the Supplemental Material). A clear exception is the case $\alpha_2 = \alpha_3$ (Fig.~\ref{fig_3}(e)), for which, in easy adaptation, $I_\text{A}$ always increases with $k^{(2)}$. The discrepancy is explained by observing that, on one hand, such increase gradually vanishes when $\rho$ approaches small values. Indeed, with a low proportion of A-type individuals in the population, the indirect mechanism weakening the N-N interactions which makes $I_\text{N}$, and consequently $I_\text{A}$, smaller, is largely compensated by the high epidemic pressure that anyway emerges via the more frequent N-type uniform groups. On the other hand, in the hard adaptation limit (see Fig.~S4 of the Supplemental Material), $I_\text{A}$ decreases rapidly for low values of $\rho$. In this limit, since there is no indirect modification of the N-N interactions, $I_\text{N}$ is not directly affected by group size, whereas $I_\text{A}$ decreases when trading $\{\text{A},\text{N},\text{N}\}$ triads for $\{\text{A},\text{N}\}$ pairs thanks to the gained bilateral protection. $\{\text{A},\text{N}\}$ and $\{\text{A},\text{N},\text{N}\}$ are thus the groups that regulate the dependence of $I_\text{A}$ on $k^{(2)}$, but also those within which the A-type individuals interact the most when $\rho$ is small. Combining the fact that in the intermediate scenario $q_\text{N}$ in a $\{\text{A},\text{N},\text{N}\}$ triad is not as high as in a $\{\text{A},\text{N}\}$ pair, thus departing from the easy adaptation limit, with the fact that by decreasing $\rho$ the effect of $k^{(2)}$ on $I_\text{A}$ gets weaker in easy adaptation but stronger in hard adaptation, results in the decrease of $I_\text{A}$ seen in Fig.~\ref{fig_3}(e) for low enough $\rho$.

Such sort of competition between easy and hard adaptation, determining what we observe in an intermediate scenario, is certainly present also for $\alpha_2 \neq \alpha_3$. However, it has no qualitative impact when $\alpha_2$ and $\alpha_3$ are not too close to each other. In such case, the considered intermediate scenario has the same qualitative behavior than the easy adaptation one.

Accordingly, for $\alpha_2 < \alpha_3$ (notice $\beta_{3,1}=1$ here), increasing $k^{(2)}$, not only makes the weakening of the N-N interactions sparser, but also increases the rate at which N-N interactions take place, yielding a rapid increase of $I_\text{N}$ (as we already know from $R$) which, in turn, pushes up $I_\text{A}$. At the same time, especially for lower $\rho$, type A benefits from the increase in homophily provided by a higher $k^{(2)}$, as it isolates it from the more infectious type N. Eventually, when $\rho$ is not too high and the increase of $I_\text{N}$ gets slower (see Fig.~\ref{fig_3}(c)), the two contributions to $I_\text{A}$ become comparable, giving the latter the nonmonotonic shape observed in Fig.~\ref{fig_3}(b). This holds whenever efficacy is high at least in one way, so that there is a substantial difference between the epidemic pressures of the two types; otherwise, homophily looses importance and $I_\text{A}$ just grows driven by $I_\text{N}$.

Finally, for $\alpha_2 > \alpha_3$ (see Fig.~\ref{fig_3}(h) and \ref{fig_3}(i)), $I_\text{A}$ mainly behaves as $I_\text{N}$. Indeed, since pairs are now less assortative than triads, type A cannot benefit from mixing to compensate for the larger exposure to type N implied by a higher $k^{(2)}$.

All in all, it is $I_\text{N}$ to be the main driver of the observed phenomenology in the easy adaptation scenario. This leads to the conclusion that, by exploiting the indirect weakening of the N-N contacts, the system (made few exceptions) benefits from interacting more in triads than in pairs. In the limit of hard adaptation that role is instead played by $I_\text{A}$ (see Sec.~S2.1 of the Supplemental Material for a detailed discussion about the results in this scenario). As observed before, while type N is not directly affected by group size in this case, type A benefits from the bilateral protection induced in $\{\text{A},\text{N}\}$ pairs, making pairs preferable to triads for the system overall. What is found for any intermediate scenario, as the one we derived from the social contagion model, eventually depends on how close this is to either one limit or the other.

\begin{figure}
    \centering
    \includegraphics[width = 1.\linewidth]{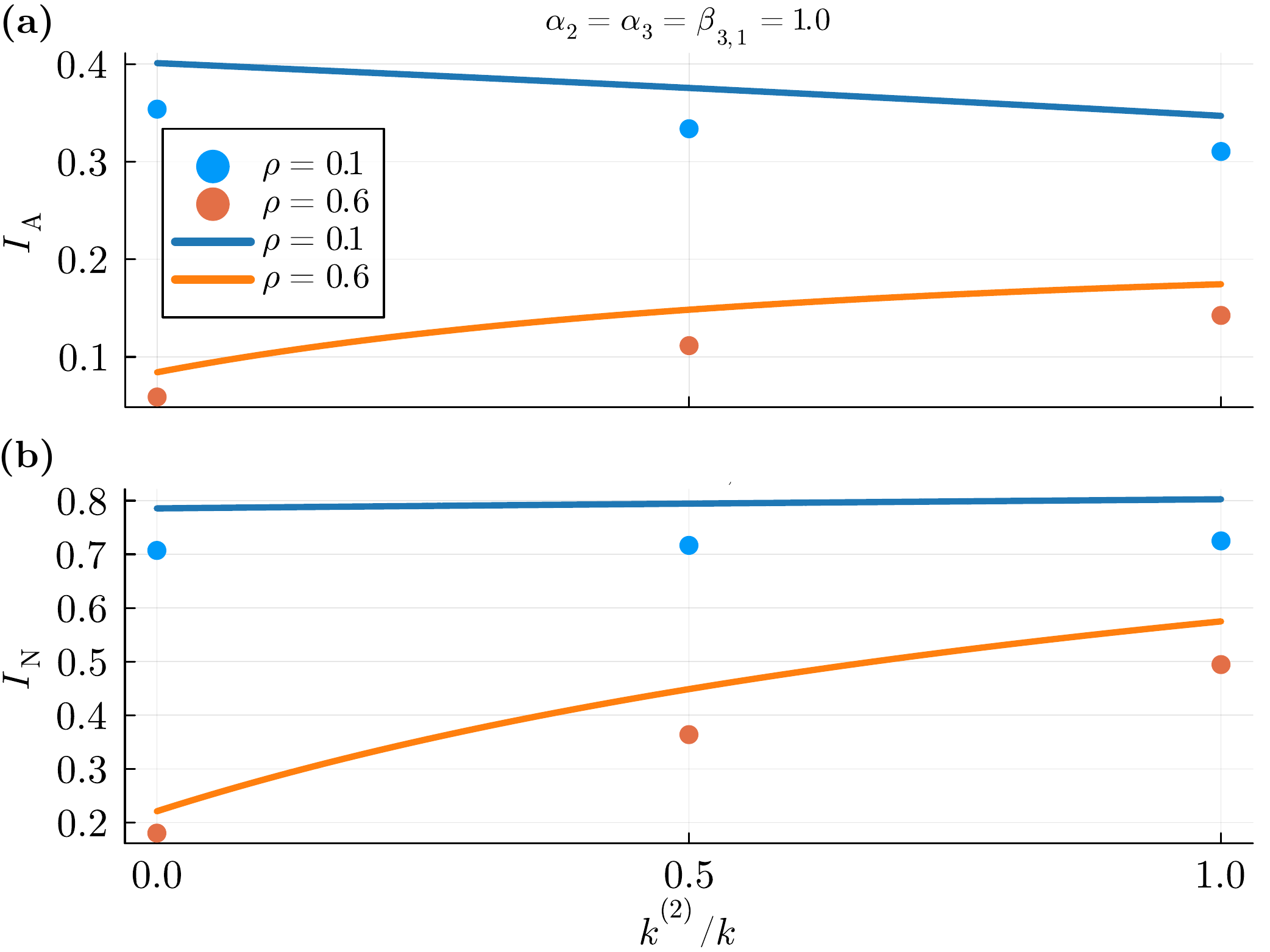}
    \caption{Monte Carlo results for the equilibrium endemic state as a function of the proportion of $2$-edges, $k^{(2)}/k$, for a regular rank-$3$ hypergraph of $N=1000$ nodes and pairwise degree $k = k^{(2)} + 2k^{(3)}= 12$. Here, $\lambda = 0.1$ , $\mu = 0.2$ and $\alpha_2 = \alpha_3 = \beta_{3,1} = 1.0$ (homogeneous mixing). {\bf (a)} Prevalence for type A, $I_\text{A}$, and {\bf (b)} for type N, $I_\text{N}$. Each point is obtained by averaging over $2500$ runs with random initial conditions (infectious state and type assignment). Error bars, representing standard deviations, are smaller than point size. Solid lines represent the results found under the mean-field approximation.}
    \label{fig_4}
\end{figure}

At last, we considered quenched contact structures (see Appendix~\ref{sec:sim} for the numerical implementation of the model) to test whether the presence of topological correlations changed the phenomenology predicted by the mean-field model. Specifically, keeping fixed the mixing parameters to some values, we varied $k^{(2)}/k$ for rank-3 hypergraphs and looked at the equilibrium endemic state. As Fig.~\ref{fig_4} shows, the qualitative behavior is accurately reproduced by the mean-field model. Note that the systematic overestimation of the prevalence is actually expected for approximation schemes that ignore dynamical correlations when the system is enough above the epidemic threshold~\cite{burgio2021network}.

\subsection{Varying the type-assortativity distribution among group sizes}
\label{sec:results_assor}

We briefly summarize here the results obtained when distributing the assortativity differently between $2$- and $3$-edges. That is, we fix $\alpha$ and vary $\alpha_2$, with the constraint $\alpha_3 = \alpha + \left(\alpha-\alpha_2\right)k^{(2)}/\left(2k^{(3)}\right)$ (Eq.~(\ref{eq:alpha3_1,0})).

In Appendix~\ref{sec:R_assor} we prove that the reproduction number increases with $\alpha_2$ in the easy adaptation scenario, whereas it decreases in the hard adaptation one.

The same holds also for each type-specific prevalence, hence for the prevalence overall (see Figs.~S3 and S5 of the Supplemental Material), and is explained as follows. In the easy adaptation limit, a higher (lower) $\alpha_2$ ($\alpha_3$) means a lower rate at which $\{\text{A},\text{N},\text{N}\}$ triads form, implying that less N-N interactions are weakened. As a consequence, the N-type's prevalence, $I_\text{N}$, increases. At the same time, since in this scenario the A-N interactions are dynamically equivalent whether they occur isolated or within a triad, the A-type's probability of infection is not directly affected by the assortativity distribution. However, since $I_\text{N}$ increases, $I_\text{A}$ increases as well (unless $\varepsilon_\text{in}=1$), although in a milder way being the A-N contacts always bilaterally protected. In hard adaptation, instead, there is no indirect mechanism modifying the N-N interactions, hence $I_\text{N}$ is not directly affected by the assortativity distribution. On the other hand, rising (lowering) $\alpha_2$ ($\alpha_3$) means increasing the rate at which the A-N interactions are bilaterally protected, implying a lower $I_\text{A}$. In any case, infection from an A-type's is unlikely (unless $\varepsilon_\text{in}$ and $\varepsilon_\text{out}$ are both small), therefore $I_\text{N}$ is only slightly reduced by $\alpha_2$.

As already observed in Sec.~\ref{sec:results_contact}, what to expect in any intermediate behavioral scenario depends then on how close it is to either one of the two limit scenarios, the results being generally an interpolation between the two (see Sec.~S3 of the Supplemental Material for the results found in the intermediate scenario used in Sec.~\ref{sec:results_contact}).

To close the analysis, note that the same qualitative effects of increasing (decreasing) $\alpha_2$ ($\alpha_3$) are also implied by increasing the higher-order mixing parameter $\beta_{3,1}$, for the frequency of the $\{\text{A},\text{N},\text{N}\}$ triads is either way reduced (see Sec.~S3 and Fig.~S6(b) of the Supplemental Material). While this should not be surprising, as $\beta_{3,1}$ determines the structure as much as any pairwise parameter ($\alpha_2$ and $\alpha_3$), it remarks the fact that exclusively relying on pairwise information may not guarantee an accurate description of the system.

\section{CONCLUSIONS}
\label{sec:conclusions}

We define a minimal model of context-dependent spreading where, during an interaction, an agent either actively behave to alter the diffusion or not depending on the behavior it observes among the co-present peers. Considering populations where agents are divided into two types (encoding different intrinsic inclinations to take on active behavior), we provide a mean-field approximation for heterogeneous mixing in hypergraphs, allowing to parametrize mixing patterns of arbitrary type-(dis)assortativity within groups of any size. Choosing then a (spreading) dynamics, the theory makes it possible to obtain analytical results that provide a basic understanding of the system.

Referring to an epidemic spreading model where context-dependency concerns the adoption of prophylactic behaviors like face-mask wearing, we have shown that accounting for the behavioral dynamics unfolding at the higher-order level of organization of the interactions, can lead to important deviations from what can be expected based on pairwise information alone. We have revealed that the direction and the magnitude of those deviations primarily depend on the properties of the behavioral dynamics; more specifically, on how effective are the (non) adoption-inclined agents in inducing (inhibiting) adoption of active behavior on the others. Then, depending also on the proportion of adoption-inclined agents in the population and the way type-assortativity distributes among the group sizes, and, secondarily, on the prophylactic efficacy, gathering more often in pairs than in triads (larger groups) can either facilitate or impede the spreading. We have proven this analytically for the basic reproduction number and shown how exclusively changing the group-size distribution can determine whether an outbreak will be subcritical ---and eventually vanish--- or supercritical ---leading to endemicity.

More specifically, either for the reproduction number and the prevalence, we can conclude that in general (made few exceptions) when prophylactic behavior is easy to induce, then the system generally benefits by interacting in larger groups, for many otherwise unprotected contacts would be now protected thanks to the elicited adoption. Moreover, the benefit increases when the smaller (larger) groups are the more (dis)assortative. On the contrary, when adoption is hard to induce (e.g., when some majority rule applies), smaller (and disassortative) groups become preferable.

Looking at each type-specific prevalence, we have seen that the type of an agent, not only ---as expected--- strongly affects its infection risk but also changes the qualitative dependence of that risk on the parameters characterizing the contact structure. Observing this becomes especially important when, for instance, adoption is mostly driven by the vulnerability to the spreading disease. For example, if elderly and young people have very different chances of suffering from severe symptoms ---as it is for COVID-19~\cite{sorensen2022variation} or influenza~\cite{cozza2021global}---, they may also have a dissimilar propensity to protect themselves and the others \cite{badillo2021global,haischer2020wearing,sanchez2021factors,kaewpan2022factors}. In such cases, the overall prevalence might not be the most useful indicator, for the prevalence within the vulnerable (and probably adoption-inclined) class does not generally follow the same phenomenology.

Our work stimulates a new direction in the discourse on the interplay between behavior and epidemic spreading. It helps to understand the role of contextual information in decisions to reduce the impact of infectious spreading and provides theoretical support for epidemic control policies based on public gathering restrictions.

In the end, the mean-field approximation provided here opens new possibilities in the modeling of higher-order systems, specifically serving as a basic theory to investigate virtually any process for which mutable and contextual factors can be relevant. Inspired by growing evidence in ecology, we have studied one of such processes and suggested a few more, but perhaps many others have not yet been recognized. We hope that other researchers will be encouraged to look for mechanisms of indirect interaction also in other systems and address the highly challenging problems their complexity entails. \\

\section*{ACKNOWLEDGMENTS}

G.B.\ acknowledges financial support from the European Union's Horizon 2020 research and innovation program under the Marie Sk\l{}odowska-Curie Grant Agreement No.\ 945413 and from the Universitat Rovira i Virgili (URV). A.A.\ and S.G.\ acknowledge support by Ministerio de Econom\'ia y Competitividad (Grants No.\ PGC2018-094754-B-C21, FIS2015-71582-C2-1 and RED2018-102518-T), Generalitat de Catalunya (Grant No.\ 2017SGR-896) and Universitat Rovira i Virgili (Grant No.\ 2019PFR-URV-B2-41). A.A.\ also acknowleges ICREA Academia, and the James S.\ McDonnell Foundation (Grant No.\ 220020325).\ We thank Clara Granell for helping with the visualization in Fig.~1.

\appendix

\section*{Appendix}

\setcounter{figure}{0}
\renewcommand\thefigure{\thesection\arabic{figure}}
\renewcommand{\theHfigure}{\thefigure}

\section{MEAN-FIELD MIXING IN GROUPS}
\label{sec:app_modeling}
\subsection{Counting the subsets of given type-composition}
\label{sec:app_modeling_g}

In this section we derive the expression for $g_{m_\text{A},m_\text{N}}^{(n)}$, given in Eqs.~(\ref{eq:g}) to~(\ref{eq:g_A}). Consider a subset $\left\{\text{A}_1,\dots,\text{A}_{m_\text{A}},\text{N}_1,\dots,\text{N}_{m_\text{N}}\right\}$ of $m_\text{A}\geqslant 1$ A-type nodes and $m_\text{N}\geqslant 1$ N-type ones. We want to calculate the probability of finding the other $m_\text{A}+m_\text{N}-1$ nodes in the subset conditioned on a single node that we can choose to be, for instance, $\text{N}_1$, that is, $P\left(\left\{\text{A}_1,\dots,\text{A}_{m_\text{A}},\text{N}_2,\dots,\text{N}_{m_\text{N}}\right\}\vert \text{N}_1\right)$. There are $(m_\text{A}+m_\text{N}-1)!$ different ways of writing this probability, one for each possible order in which the $m_\text{A}+m_\text{N}-1$ nodes can be found. We can thus choose one particular ordering, say $\left(\text{N}_2,\dots,\text{N}_{m_\text{N}},\text{A}_1,\dots,\text{A}_{m_\text{A}}\right)$, and apply recursively the definition of conditional probability (i.e., $P(\{x,y\})=P(y)P(x\vert y)$ and $P(\{x,y\}\vert z)=P(y\vert z)P(x\vert \{y,z\}$) to get

\noindent
\begin{align}
    \notag
    P(\{&\text{A}_1,\dots,\text{A}_{m_\text{A}},\text{N}_2,\dots,\text{N}_{m_\text{N}}\}\vert \text{N}_1)
    \\ \notag =&
    \ (m_\text{A}+m_\text{N}-1)!
    \\ \notag & \times
    P\left(\text{N}_2\vert \text{N}_1\right)
    \\ \notag &
    \ \ \ \ \vdots
    \\ \notag & \times
    P\left(\text{N}_{m_\text{N}}\vert \left\{\text{N}_1,\dots,\text{N}_{m_\text{N}-1}\right\}\right)
    \\ \notag & \times
    P\left(\text{A}_1\vert \left\{\text{N}_1,\dots,\text{N}_{m_\text{N}}\right\}\right)
    \\ \notag &
    \ \ \ \ \vdots
    \\        & \times P\left(\text{A}_{m_\text{A}}\vert \left\{\text{N}_1,\dots,\text{N}_{m_\text{N}},\text{A}_1,\dots,\text{A}_{m_\text{A}-1}\right\}\right)\ .
    \label{eq:app_P_labs}
\end{align}

\noindent If we are not interested on the label of the nodes but exclusively on their type, using the notation $m\text{X}$ to denote a set of $m$ X-type nodes, Eq.~(\ref{eq:app_P_labs}) becomes
\begin{align}
    \notag P(&m_\text{A}\text{A},\left(m_\text{N}-1\right)\text{N}\vert \text{N})
    \\ \notag =&\ \binom{m_\text{A}+m_\text{N}-1}{m_\text{A}}
    \\ \notag & \times
    P\left(\text{N}\vert \text{N}\right) P\left(\text{N}\vert 2\text{N}\right) \cdots P\left(\text{N}\vert \left(m_\text{N}-1\right)\text{N}\right) \\        & \times
    P\left(\text{A}\vert m_\text{N}\text{N}\right) \cdots P\left(\text{A}\vert m_\text{N}\text{N},\left(m_\text{A}-1\right)\text{A}\right)\ ,
    \label{eq:app_P_types}
\end{align}

\noindent where we divided by $\left(m_\text{N}-1\right)!$ and $m_\text{A}!$, which are the number of permutations of the $m_\text{N}-1$ N-type and $m_\text{A}$ A-type nodes in the sequence, respectively, equally contributing to $P(m_\text{A}\text{A},\left(m_\text{N}-1\right)\text{N}\vert \text{N})$.

If the $m_\text{A}+m_\text{N}-1$ nodes are drawn from a set of $n-1 \geqslant m_\text{A}+m_\text{N}-1$ nodes, there are $\binom{n-1}{m_\text{A}+m_\text{N}-1}$ distinct ways of doing it for any sequence chosen for the $m_\text{A}+m_\text{N}-1$ nodes. Then, if there is a total of $N_\text{N}$ N-type nodes in the population and, on average, each one is included in $k_\text{N}^{(n)}$ $n$-edges per time step, the expected number $g_{m_\text{A},m_\text{N}}^{(n)}$ of $\left(m_\text{A}+m_\text{N}\right)$-edges composed of $m_\text{A}$ A-type nodes and $m_\text{N}$ N-type ones, subsets of $n$-edges, can be written as
\begin{align}
    \notag g_{m_\text{A},m_\text{N}}^{(n)} =& \frac{N_\text{N} k_\text{N}^{(n)}}{m_\text{N}} \binom{n-1}{m_\text{A}+m_\text{N}-1} \binom{m_\text{A}+m_\text{N}-1}{m_\text{A}} \\
    \notag & \times ~ P\left(\text{N}\vert \text{N}\right) P\left(\text{N}\vert 2\text{N}\right) \cdots P\left(\text{N}\vert \left(m_\text{N}-1\right)\text{N}\right) \\
    &\times ~ P\left(\text{A}\vert m_\text{N}\text{N}\right) \cdots P\left(\text{A}\vert m_\text{N}\text{N},\left(m_\text{A}-1\right)\text{A}\right)\ ,
    \label{eq:app_g}
\end{align}

\noindent where the term $1/m_\text{N}$ accounts for the fact that there are $m_\text{N}$ N-type nodes to condition upon in the set. If the subset is N-type uniform, we can just take $m_\text{A}=0$ in Eq.~(\ref{eq:app_g}) to obtain
\begin{align}
    \notag g_{0,m_\text{N}}^{(n)} =& \frac{N_\text{N} k_\text{N}^{(n)}}{m_\text{N}} \binom{n-1}{m_\text{N}-1}
    \\ &\times
    P\left(\text{N}\vert \text{N}\right) P\left(\text{N}\vert 2\text{N}\right) \cdots P\left(\text{N}\vert \left(m_\text{N}-1\right)\text{N}\right)\ .
    \label{eq:app_g_uni}
\end{align}
\noindent Analogous expressions are found by initially conditioning on an A-type. Apart from the more explicit notation used here, Eqs.~(\ref{eq:app_g}) and~(\ref{eq:app_g_uni}) are identical to Eqs.~(\ref{eq:g}) and~(\ref{eq:g_N}).

\subsection{Number of free parameters}
\label{sec:app_modeling_params}

Since $m_\text{A}\in\{0,\dots,m\}$, Eqs.~(\ref{eq:pn1}) and (\ref{eq:pn2}) imply that there are $m+1$ parameters characterizing the mixing within $(m+1)$-edges. To see that only one of them is free, observe from Eq.~(\ref{eq:pn}) that the number of mixed groups $g_{m_\text{A}+l_\text{A},m_\text{N}+l_\text{N}}^{(n)}$ ($m_\text{A}$, $m_\text{N}\geqslant 1$) can be expressed in terms of both $p_{l_\text{A},l_\text{N}\vert m_\text{A},m_\text{N}}^{(n)}$ and $p_{l_\text{A}-1,l_\text{N}+1\vert m_\text{A}+1,m_\text{N}-1}^{(n)}$ if $l_\text{A} = 1$ ($l_\text{N} = 0$), establishing a relation between $\alpha_{m,m_\text{A}}^{(n)}$ and $\alpha_{m,m_\text{A}+1}^{(n)}$; or, alternatively, in terms of both $p_{l_\text{A},l_\text{N}\vert m_\text{A},m_\text{N}}^{(n)}$ and $p_{l_\text{A}+1,l_\text{N}-1\vert m_\text{A}-1,m_\text{N}+1}^{(n)}$ if $l_\text{N} = 1$ ($l_\text{A} = 0$), providing a relation between $\alpha_{m,m_\text{A}}^{(n)}$ and $\alpha_{m,m_\text{A}-1}^{(n)}$. Since there are $m$ distinct mixed group configurations (one for each $m_\text{A}\in\{1,\dots,m\}$), there are $m$ constraints relating the $m+1$ mixing parameters, which can thus be expressed in terms of only one of them. Those relations can be found by simply substituting Eqs.~(\ref{eq:pn1}) and~(\ref{eq:pn2}) into Eqs.~(\ref{eq:g}) to~(\ref{eq:g_A}) and, as explained before, comparing related pairs of conditional probabilities (or, equivalently, expressing $g_{m_\text{A},m_\text{N}}^{(n)}$ using different orders of search). Notice also that the sets $\{\alpha_{m,m_\text{A}}^{(n)}\}$ and $\{\alpha_{m,m_\text{A}}^{(n^\prime)}\}$, $m < \text{min}\left\{n,n^\prime\right\}$, satisfy the same set of relations, as the form of Eqs.~(\ref{eq:pn1}) and~(\ref{eq:pn2}) does not depend on $m$: $\alpha_{m,m_\text{A}}^{(n)}$ and $\alpha_{m,m_\text{A}}^{(n^\prime)}$ can only differ is their value, being they computed on different sets ($n$- and $n^\prime$-edges, respectively). Therefore, we can find all the relations by just considering the case $m=n-1$ for each $n\in\{2,\dots,n_\text{max}\}$. Having one free parameter for each $m\in\{1,\dots,n-1\}$, we get $n-1$ parameters to fix for group size $n$. Being $n\in\{2,\dots,n_\text{max}\}$, the interaction structure is then determined by at most $\sum_{n=2}^{n_\text{max}} (n-1) = \binom{n_\text{max}}{2}$ parameters.

\section{DERIVATION OF THE SOCIAL CONTAGION MODEL}
\label{sec:behav_dyn}

We detail here the derivation of the social contagion model presented in Sec.~\ref{sec:social contagion}. Consider a group of size $n$, composed of $n_\text{A}$ and $n_\text{N}$ agents of type A and N, respectively, and denote with $q_\text{X}(t)$ the probability of adopting an active behavior for a X-type's at time $t$. An A-type agent in the adopter state is induced (e.g., through peer pressure) to switch to the nonadopter state by the currently nonadopters in the group. These are, on average, $\left(1-q_\text{A}(t)\right)\left(n_\text{A}-1\right) + \left(1-q_\text{N}(t)\right)n_\text{N}$. If, instead, the A-type's is in the nonadopter state, it is induced to switch to the adopter one by the $q_\text{A}(t)\left(n_\text{A}-1\right) + q_\text{N}(t)n_\text{N}$ adopters that, on average, it currently finds in the group. The same holds for a N-type agent given $n_\text{A}-1$ and $n_\text{N}$ in the expressions above are replaced by $n_\text{A}$ and $n_\text{N}-1$, respectively. Additionally, agents can spontaneously move between the two behavioral compartments at rates depending solely on their type. These rates quantify some cost of adopting an active behavior and the will to bear that cost. Denoting with $c_\text{X}$ the I-to-S rate and with $b_\text{X}$ the S-to-I rate, for type X, the adaptive behavioral dynamics is described by the following system of two differential equations
\begin{align}
    \notag \dot{q_{\text{A}}} =& \left(1-q_\text{A}\right)\left[q_\text{A}\frac{n_\text{A}-1}{n-1} + q_\text{N}\frac{n_\text{N}}{n-1}\right] \\
    \notag &- q_\text{A}\left[\left(1-q_\text{A}\right)\frac{n_\text{A}-1}{n-1} + \left(1-q_\text{N}\right)\frac{n_\text{N}}{n-1}\right] \\
    &+ b_{\text{A}}\left(1-q_{\text{A}}\right) - c_{\text{A}}q_{\text{A}}\ , \label{eq:qA_} \\
    \notag \dot{q_{\text{N}}} =& \left(1-q_\text{N}\right)\left[q_\text{N}\frac{n_\text{N}-1}{n-1} + q_\text{A}\frac{n_\text{A}}{n-1}\right]\\
    \notag &- q_\text{N}\left[\left(1-q_\text{N}\right)\frac{n_\text{N}-1}{n-1} + \left(1-q_\text{A}\right)\frac{n_\text{A}}{n-1}\right] \\
    &+ b_{\text{N}}\left(1-q_{\text{N}}\right) - c_{\text{N}}q_{\text{N}}\ , \label{eq:qN_}
\end{align}

\noindent where $q_{\text{X}}\equiv q_{\text{X}}^{(n_\text{X}-1,n-1)}(t)$ for a lighter notation. The $n-1$ in the denominator comes from assuming that each of the $n-1$ sources weights the same in affecting the behavior of a focal agent. It is immediate to see that the terms involving only one type ($\propto\left(1-q_\text{X}\right)q_\text{X}$), as well as the mixed quadratic terms ($\propto q_\text{A} q_\text{N}$), cancel out. Therefore, we are left with the linear system defined by Eqs.~(\ref{eq:qA}) and (\ref{eq:qN}), and the solutions described after it.

In the solution for mixed groups (i.e., $1\leqslant n_\text{A}\leqslant n-1$), the constants $C_1$ and $C_2$, which depend on all the figuring parameters, read
\begin{align}
    C_1 &= \frac{\left(n-1\right)b_\text{A}b_\text{N} + \ds\sum_{\text{X}=\text{A},\text{N}}n_\text{X}b_\text{X}}{\left(n-1\right)\left(b_\text{A}+c_\text{A}\right)\left(b_\text{N}+c_\text{N}\right) + \ds\sum_{\text{X}=\text{A},\text{N}} n_\text{X}\left(b_\text{X}+c_\text{X}\right)}\ , \label{eq:C1} \\
    C_2 &= \frac{n-1}{\left(n-1\right)\left(b_\text{A}+c_\text{A}\right)\left(b_\text{N}+c_\text{N}\right) + \ds\sum_{\text{X}=\text{A},\text{N}} n_\text{X}\left(b_\text{X}+c_\text{X}\right)}\ . \label{eq:C2}
\end{align}

\noindent The difference $q_{\text{A}}-q_{\text{N}}=C_2\left(b_{\text{A}}c_{\text{N}} - b_{\text{N}}c_{\text{A}}\right)$ requires $b_{\text{A}}c_{\text{N}}>b_{\text{N}}c_{\text{A}}$ in order for A-type agents to be actually more inclined to be adopters than N-type's are. We can thus choose $b_{\text{A}} > c_{\text{A}}$ (or $b_{\text{A}} \gg c_{\text{A}}$) and $b_{\text{N}} < c_{\text{N}}$ (or $b_{\text{N}} \ll c_{\text{N}}$). Also, it requires $C_2$ to be large enough and, from Eq.~(\ref{eq:C2}), we see this means considering not too large values for the rates (specifically for $b_{\text{A}}$ and $c_{\text{N}}$, when $b_{\text{A}} \gg c_{\text{A}}$ and $b_{\text{N}} \ll c_{\text{N}}$). The equilibrium adoption probability for the two types as a function of the composition of a group for different values of the A-type's adoption rate, $b_\text{A}$, is shown in Sec.~S1 of the Supplemental Material for groups of $n=2,3,4$ individuals, making evident the nonlinear dependence on the composition. The nonlinearity increases with $b_\text{A}$ and tends to disappear only for $b_\text{A}$ approaching $c_\text{A}$.

\section{DEPENDENCE OF THE BASIC REPRODUCTION NUMBER ON THE STRUCTURAL PARAMETERS}
\label{sec:R_par}

\subsection{Degree}
\label{sec:R_contact}

The basic reproduction number, $R=\lambda k_{\text{eff}}/\mu$, is here studied as a function of the $2$-degree, $k^{(2)}$, while keeping fixed the total pairwise degree, $k=k^{(2)}+2k^{(3)}$. We assume type and degree to be uncorrelated, i.e., $k_\text{A}^{(n)}=k_\text{N}^{(n)}=k^{(n)}$. For any fixed $k$, we want to establish the sign of $\partial R/\partial k^{(2)}\vert_{k} = \left(\lambda/\mu\right)\partial{k_{\text{eff}}}/\partial k^{(2)}\vert_{k}$. To this end, given the complicated expression of $k_{\text{eff}}$ (Eq.~(\ref{eq:k_eff})), we simplify it by referring to the binary behavioral scenarios presented in Sec.~\ref{sec:binary}. In both of them, it holds $r_\text{A}=1-\varepsilon_\text{out}$ and $s_\text{A}=1-\varepsilon_\text{in}$. Moreover, we assume perfect protection in at least one way, that is, $\varepsilon_\text{in}=1$ and/or $\varepsilon_\text{out}=1$. \\

\paragraph{Easy adaptation.}

Here, $r_\text{N}=1-\varepsilon_\text{out}$ and $s_\text{N}=1-\varepsilon_\text{in}$ for $1\leqslant n_\text{N}\leqslant n-1$, with $\varepsilon_\text{in}=1$ and/or $\varepsilon_\text{out}=1$, and  $r_\text{N} = s_\text{N}= 1$ for $n_\text{N} = n$. Only $\theta_{\text{N}\rightarrow\text{N}}$ is nonzero, hence $k_{\text{eff}} = \theta_{\text{N}\rightarrow\text{N}}$ and is given by
\begin{equation}
    k_{\text{eff}} = k^{(2)} \left(1 - \alpha_2 \rho\right) + \left(k-k^{(2)}\right) \left[1 - \alpha_3 \rho \left(2 - \beta_{3,1} \rho\right)\right]\ ,
\end{equation}

\noindent where the two terms multiplying the degrees account for the probability that a N-type individual takes part to a type-N uniform group of size $2$ and $3$, respectively, for these are the groups where it is not an adopter. Then we obtain
\begin{equation}
    \left.\frac{\partial k_{\text{eff}}}{\partial k^{(2)}}\right\vert_{k} = \rho\left[\alpha_3 \left(2 - \beta_{3,1} \rho\right)-\alpha_2\right]\ ,
    \label{eq:dR/dk2}
\end{equation}

\noindent and equaling it to zero, we get the solutions $\rho=0$ and $\rho=\tilde\rho$, with
\begin{equation}
    \tilde\rho = \frac{2\alpha_3 - \alpha_2}{\alpha_3\beta_{3,1}}\ ,
    \label{eq:rho_th}
\end{equation}

\noindent given $\alpha_3\beta_{3,1}\neq 0$. Also, note that $\tilde\rho\in\left[0,1\right]$ requires $\alpha_3\in\left[\alpha_2/2,\alpha_2/\left(2-\beta_{3,1}\right)\right]$. Technically, $\tilde\rho$ results from the fact that the frequency of the \{A,A,N\} and \{A,N,N\} triads are both quadratic (respectively, increasing and decreasing) functions of $\rho^2$, whereas the frequency of the \{A,N\} pairs increases linearly with $\rho$. Finally, differentiating Eq.~(\ref{eq:dR/dk2}) with respect to $\rho$ and computing it at $\rho = 0$ and $\rho = \tilde\rho$, one finds
\begin{equation}
    \frac{\partial}{\partial \rho}\left.\frac{\partial k_{\text{eff}}}{\partial k^{(2)}}\right\vert_{k} =
    \begin{cases}
      2\alpha_3-\alpha_2 & \text{at $\rho = 0$}\ , \\
      \alpha_2-2\alpha_3 & \text{at $\rho = \tilde\rho$}\ .
    \end{cases}
    \label{eq:d^2R/drhodk2}
\end{equation}

\noindent Thus, given $\tilde\rho>0$ (i.e., $\alpha_3>\alpha_2/2$), it follows that $\partial k_{\text{eff}}/\partial k^{(2)}\vert_{k}$ is positive for $\rho\in\left(0,\tilde\rho\right)$ and negative for $\rho\in\left(\tilde\rho,1\right]$; otherwise, $\partial k_{\text{eff}}/\partial k^{(2)}\vert_{k}<0$ for any $\rho\in\left(0,1\right]$. Rephrasing it, $k_{\text{eff}}$ (hence $R$) is an increasing or decreasing function of $k^{(2)}$ depending on whether $\rho$ is lower or higher than $\tilde \rho$. \\

\paragraph{Hard adaptation.}

In this scenario, $r_\text{N}=1-\varepsilon_\text{out}$ and $s_\text{N}=1-\varepsilon_\text{in}$ for $n_\text{N}=1$, with $\varepsilon_\text{in}=1$ and/or $\varepsilon_\text{out}=1$, and  $r_\text{N} = s_\text{N}= 1$ otherwise. We get
\begin{equation}
    k_{\text{eff}} = k^{(2)} \left(1 - \alpha_2 \rho\right) + \left(k-k^{(2)}\right) \left(1 - \alpha_3 \rho\right)\ ,
\end{equation}

\noindent the two terms ---notice, both linear in $\rho$--- accounting for the probability that a N-type individual finds at least another N-type's in a group of size $2$ and $3$, respectively, not being an adopter in such cases. We obtain
\begin{equation}
    \left.\frac{\partial k_{\text{eff}}}{\partial k^{(2)}}\right\vert_{k} = \rho\left(\alpha_3 - \alpha_2\right)\ ,
    \label{eq:dR/dk2_}
\end{equation}

\noindent therefore $k_{\text{eff}}$ ($R$) increases or decreases with $k^{(2)}$ solely depending on whether $\alpha_3$ is larger or smaller than $\alpha_2$, respectively.

\subsection{Assortativity}
\label{sec:R_assor}

Here we study how the basic reproduction number is affected by the way in which assortativity is distributed between $2$- and $3$-edges. No correlation is assumed between type and degree. Fixed the average pairwise mixing parameter, $\alpha = \left(\alpha_2 k^{(2)} + 2\alpha_3 k^{(3)}\right)/k$, with $k=k^{(2)}+2k^{(3)}$, we want to establish the sign of $\partial{R}/\partial{\alpha_2}\vert_{\alpha}=\left(\lambda/\mu\right)\partial{k_{\text{eff}}}/\partial{\alpha_2}\vert_{\alpha}$. \\

\paragraph{Easy adaptation.}

Here, $r_\text{N}=1-\varepsilon_\text{out}$ and $s_\text{N}=1-\varepsilon_\text{in}$ for $1\leqslant n_\text{N}\leqslant n-1$, and  $r_\text{N} = s_\text{N}= 1$ for $n_\text{N} = n$. After some algebra, one gets to
\begin{align}
    \notag \left.\frac{\partial{k_{\text{eff}}}}{\partial{\alpha_2}}\right\vert_{\alpha} &= \ \frac{k^{(2)}}{2}\rho\left(1-\beta_{3,1}\rho\right)\left(\varepsilon_\text{out}+\varepsilon_\text{in}-\varepsilon_\text{out}\varepsilon_\text{in}\right) \\
    \times &\frac{C\ + \left[C^2 + \left[2k\alpha\left(1-\varepsilon_\text{out}\right)\left(1-\varepsilon_\text{in}\right)\right]^2\rho\left(1-\rho\right)\right]^{\frac12}}{\left[C^2 + \left[2k\alpha\left(1-\varepsilon_\text{out}\right)\left(1-\varepsilon_\text{in}\right)\right]^2\rho\left(1-\rho\right)\right]^{\frac12}}
    \label{eq:der_k_eff_1} \\
    \notag \geqslant &\ \frac{k^{(2)}}{2}\rho\left(1-\beta_{3,1}\rho\right)\left(\varepsilon_\text{out}+\varepsilon_\text{in}-\varepsilon_\text{out}\varepsilon_\text{in}\right) \\
    \times &\frac{C\ + \vert C\vert}{\left[C^2 + \left[2k\alpha\left(1-\varepsilon_\text{out}\right)\left(1-\varepsilon_\text{in}\right)\right]^2\rho\left(1-\rho\right)\right]^{\frac12}} \geqslant 0\ .
    \label{eq:der_k_eff_1_ineq}
\end{align}

\noindent Whatever the value of $C$, which is a function of the various parameters, since the first term in Eq.~(\ref{eq:der_k_eff_1}) is nonnegative (recall the condition $\beta_{3,1}\leqslant 1/\rho$, Eq.~(\ref{eq:lim3})) and the second term within the square root as well, immediately follow both inequalities. From Eq.~(\ref{eq:der_k_eff_1}) we see that $\partial{k_{\text{eff}}}/\partial{\alpha_2}\vert_{\alpha}$ is strictly positive when at least one between $\varepsilon_\text{out}$ and $\varepsilon_\text{in}$ is nonzero, i.e., trivially, whenever there is protection. Additionally, we note that the dependence of $k_{\text{eff}}$ on $\alpha_2$ is reduced by making $1-\beta_{3,1}\rho$ smaller. From Eqs.~(\ref{eq:p3_4}) and (\ref{eq:p'3_2}) we see that that means lowering the frequency of $\{\text{A},\text{N},\text{N}\}$ triads, confirming them as the responsible for that dependence. \\

\paragraph{Hard adaptation.}

In this scenario, $r_\text{N}=1-\varepsilon_\text{out}$ and $s_\text{N}=1-\varepsilon_\text{in}$ for $n_\text{N}=1$, and  $r_\text{N} = s_\text{N}= 1$ otherwise. With a bit of algebra, one finds
\begin{align}
    \notag \left.\frac{\partial k_{\text{eff}}}{\partial{\alpha_2}}\right\vert_{\alpha} \propto & -k^{(2)} \rho\left(1-\rho\right)\left(1-\beta_{3,1}\rho\right)\left(1-\varepsilon_\text{out}\right)\left(1-\varepsilon_\text{in}\right) \\
    &\times \notag \left[\varepsilon_\text{out}+\varepsilon_\text{in}-2\varepsilon_\text{out}\varepsilon_\text{in}\right. \\
    &\ \ \ \ \left.\times\left(\beta_{3,1}\rho + \left(1-\beta_{3,1}\rho\right)\frac{\alpha_2 k^{(2)}}{\alpha k}\right)\right] \alpha k \label{eq:der_k_eff_2} \\
    \notag \leqslant & -k^{(2)} \rho\left(1-\rho\right)\left(1-\beta_{3,1}\rho\right)\left(1-\varepsilon_\text{out}\right)\left(1-\varepsilon_\text{in}\right)\\
    &\times \left(\varepsilon_\text{out}+\varepsilon_\text{in}-2\varepsilon_\text{out}\varepsilon_\text{in}\right) \alpha k \leqslant 0\ ,
    \label{eq:der_k_eff_2_ineq}
\end{align}

\noindent where the constant of proportionality in Eq.~(\ref{eq:der_k_eff_2}) is positive. The first inequality comes from the constraints $\alpha_2\leqslant \alpha k/k^{(2)}$, ensuring $\alpha_3\geqslant 0$ (combining Eqs.~(\ref{eq:lim2}) and (\ref{eq:alpha3_1,0})), and $\beta_{3,1}\leqslant 1/\rho$ (Eq.~(\ref{eq:lim3})), so that the term in square brackets is a decreasing function of $\alpha_2$ and thus takes its minimum value at $\alpha_2= \alpha k/k^{(2)}$. This yields the first inequality in Eq.~(\ref{eq:der_k_eff_2_ineq}) and, since the remaining terms are all nonnegative, we finally have $\partial{k_{\text{eff}}}/\partial{\alpha_2}\vert_{\alpha}\leqslant 0$. Comparing with the previous case, we see that the sign of $\partial{k_{\text{eff}}}/\partial{\alpha_2}\vert_{\alpha}$ is thus dictated by whether the adaptation by the N-type individuals is easy or hard. From Eq.~(\ref{eq:der_k_eff_2}) we note that, in this scenario, the derivative goes to zero for either full out-going ($\varepsilon_\text{out}=1$) or in-going ($\varepsilon_\text{in}=1$) protection (besides $\varepsilon_\text{out}=\varepsilon_\text{in}=0$). In such cases, secondary infections are exclusively generated by N-types', but in absence of indirect modifications, their state is unaffected by how assortativity is distributed among group sizes. In the end, as before, lowering $1-\beta_{3,1}\rho$ reduces the dependence of $k_{\text{eff}}$ on $\alpha_2$, confirming the role of the $\{\text{A},\text{N},\text{N}\}$ triads.

\section{APPROXIMATE DYNAMICS FOR HIGH PROPHYLACTIC EFFICACY}
\label{sec:prev_approx}

Let us assume that the probability of adoption is low in type-N uniform groups and high for type A in any group. Considering first the limit of high out-going prophylactic efficacy, i.e., $1-\varepsilon_\text{out}\ll 1$, we have $\theta_{\text{A}\rightarrow\text{A}}, \theta_{\text{A}\rightarrow\text{N}} \sim O\left(1-\varepsilon_\text{out}\right)$, hence $\theta_{\text{A}\rightarrow\text{A}}\ll\theta_{\text{N}\rightarrow\text{A}}$ and $\theta_{\text{A}\rightarrow\text{N}}\ll\theta_{\text{N}\rightarrow\text{N}}$. In such regime, Eqs.~(\ref{eq:dynA}) and (\ref{eq:dynN}) approximate to
\begin{align}
    \dot{I_{\text{A}}} \approx -\mu I_{\text{A}} + \lambda \left(1-I_{\text{A}}\right) I_{\text{N}}\theta_{\text{N}\rightarrow\text{A}}\ , \label{eq:dynA_approx}
    \\
    \dot{I_{\text{N}}} \approx -\mu I_{\text{N}} + \lambda \left(1-I_{\text{N}}\right) I_{\text{N}}\theta_{\text{N}\rightarrow\text{N}}\ . \label{eq:dynN_approx}
\end{align}

\noindent Equation~(\ref{eq:dynN_approx}) is a standard SIS dynamics for $I_{\text{N}}$, hence its nonzero fixed point (stable for $\lambda\theta_{\text{N}\rightarrow\text{N}}\geqslant\mu$) reads
\begin{equation}
    {I_{\text{N}}}^\star = 1 - \frac{\mu}{\lambda\theta_{\text{N}\rightarrow\text{N}}} \approx 1 - \frac{1}{R}\ , \label{eq:IN_approx}
\end{equation}

\noindent where the approximation comes from $k_\text{eff}\approx\theta_{\text{N}\rightarrow\text{N}}$ (see Eq.~(\ref{eq:k_eff})). As a consequence, ${I_{\text{N}}}^\star$ and $R$ share the same dependence on the various parameters. The nonzero fixed point for $I_{\text{A}}$ is then
\begin{equation}
    {I_{\text{A}}}^\star = \frac{1}{1 + \frac{\mu}{\lambda\theta_{\text{N}\rightarrow\text{A}}{I_{\text{N}}}^\star}} \approx \frac{\lambda}{\mu}\theta_{\text{N}\rightarrow\text{A}} . \label{eq:IA_approx}
\end{equation}

\noindent In particular, if also the in-going efficacy is high, then Eq.~(\ref{eq:IA_approx}) implies $I_{\text{A}}^\star \ll I_{\text{N}}^\star$. From Eqs.~(\ref{eq:IN_approx}) and (\ref{eq:IA_approx}) is easy to see how ${I_{\text{N}}}^\star$ and ${I_{\text{A}}}^\star$ can show different shapes. For instance, everything else left unchanged, an overall increased homophily yields a higher $\theta_{\text{N}\rightarrow\text{N}}$ but a lower $\theta_{\text{N}\rightarrow\text{A}}$, therefore ${I_{\text{N}}}^\star$ grows (as $R$ does) but ${I_{\text{A}}}^\star$ can either increase or decrease depending on whether the rate at which $\theta_{\text{N}\rightarrow\text{A}}$ decreases is lower or higher than that at which ${I_{\text{N}}}^\star$ (growing with $\theta_{\text{N}\rightarrow\text{N}}$) increases. In other words, the reproduction number may not be informative about the endemic equilibrium for type A---not even qualitatively---, although it generally may for type N.

Considering now the limit of high in-going prophylactic efficacy, i.e., $1-\varepsilon_\text{in}\ll 1$, we get $\theta_{\text{A}\rightarrow\text{A}}, \theta_{\text{N}\rightarrow\text{A}} \sim O\left(1-\varepsilon_\text{in}\right)$, implying that $I_{\text{A}}^\star \sim O\left(1-\varepsilon_\text{in}\right)$ is the only solution. In turn, the latter yields $I_{\text{N}}^\star$ as given---again---by Eq.~(\ref{eq:IN_approx}), the solution of Eq.~(\ref{eq:dynN_approx}).

\begin{figure}
    \centering
    \includegraphics[width = 1.\linewidth]{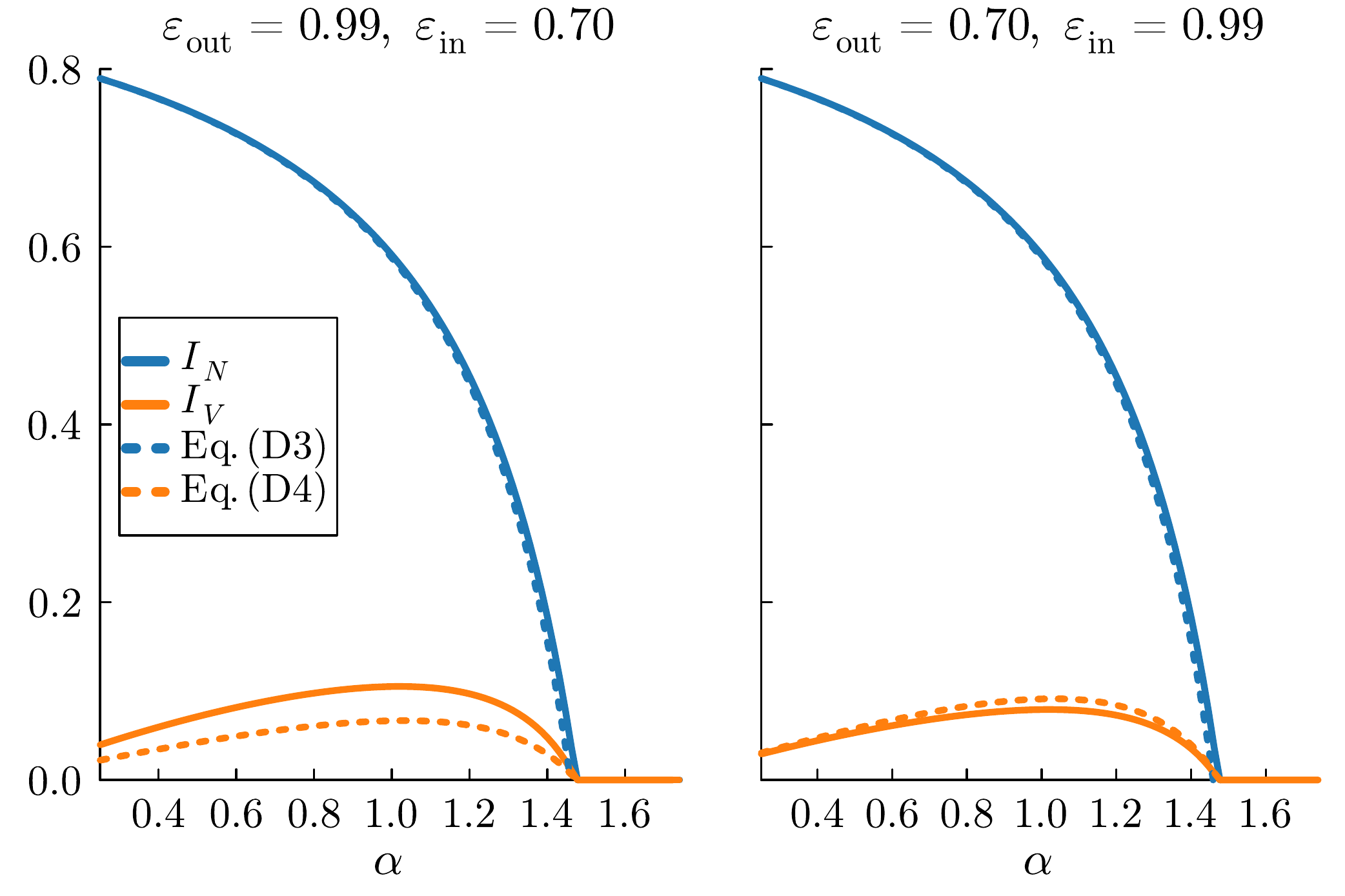}
    \caption{Exact versus approximated prevalence as given by Eqs.~(\ref{eq:IN_approx}) and (\ref{eq:IA_approx}) (dotted). Here, $\alpha_2 = \alpha_3 = \alpha$, $\beta_{3,1} = 1.0$, $\rho = 0.45$, $k = k^{(2)} + 2k^{(3)}= 6$ and $\lambda / \mu = 1.0$.}
    \label{fig_app}
\end{figure}

Figure \ref{fig_app} gives an example of how the approximation performs
for either high out-going or in-going efficacy. It is less accurate in the former case for type A because, not being the in-going efficacy high enough, the A-to-A infection rate is not so small as assumed. In any case, the approximation preserves the qualitative behavior.

\section{NUMERICAL SIMULATION OF THE MICROSCOPIC SPREADING DYNAMICS}
\label{sec:sim}

The microscopic SIS spreading dynamics is simulated as a discrete-time Markov process. Letting $\Delta t\ll 1$ be the time step duration, the local state of the nodes is updated synchronously at the discrete times $t=0,\Delta t, 2 \Delta t,\dots$, up to a maximum time step. At each step, every node interacts within each of the $k^{(n)}_i$ $n$-edges including it. The contact structure is encoded in the adjacency tensors $\left\{A^{(n)}\right\}$ such that $A_{i_1,\dots,i_n}^{(n)} = 1$ if nodes $i_1,\dots,i_n$ form a $n$-edge and $0$ otherwise. In the following we consider $n=2,3$. Denoting with $\sigma_i(t)\in\{0,1\}$ the binary variable representing the state of node $i\in\left\{1,\dots,N\right\}$ at time $t$ ($\sigma_i(t)=0$ if susceptible, $\sigma_i(t)=1$ if infected), we need to compute the transition probabilities of infection, $P\left(\sigma_i(t+\Delta t) = 1\vert \sigma_i(t) = 0\right)$, and recovery, $P\left(\sigma_i(t+\Delta t) = 0\vert \sigma_i(t) = 1\right)$. Finally, let $\text{X}_i\in\left\{\text{A},\text{N}\right\}$ be the type of node $i$.

If the interaction occurs in a $2$-edge, we have four possible values for the transmission probability, depending on the types of the two involved nodes, say $i$ and $j$. If $i$ is the susceptible node and $j$ the infected one, then the transmission rate is $\lambda r_{\text{X}_j}^{(1,1)} s_{\text{X}_i}^{(1,1)}$ if $\text{X}_i = \text{X}_j$ and $\lambda r_{\text{X}_j}^{(0,1)} s_{\text{X}_i}^{(0,1)}$ if $\text{X}_i \neq \text{X}_j$. Accordingly, the transmission probability from $j$ to $i$ during the time interval $\left[t,t+\Delta t\right]$ is, respectively, $\lambda r_{\text{X}_j}^{(1,1)} s_{\text{X}_i}^{(1,1)}\Delta t$ and $\lambda r_{\text{X}_j}^{(0,1)} s_{\text{X}_i}^{(0,1)}\Delta t$. The out-going and in-going probabilities, $r$ and $s$, are given in Eqs.~(\ref{eq:r_out}) and (\ref{eq:r_in}) and are computed at the equilibrium of the adoption dynamics (i.e., using the fixed point of Eqs.~(\ref{eq:qA})-(\ref{eq:qN})). The probability $w_i^{(2)}(t)$ that $i$ does not get infected via any of the $2$-edges incident on it, thus reads
\begin{equation}
    w_i^{(2)}(t) = \prod_{j=1}^N A_{i,j}^{(2)}\left[1 - \sigma_j(t)\lambda \tau_{i,j}^{(2)}\Delta t\right]\ ,
    \label{eq:no_inf_2}
\end{equation}

\noindent where
\begin{align}
    \notag \tau_{i,j}^{(2)} \equiv & \ \tau^{(2)}\left(\text{X}_i,\text{X}_j\right) \\
    =& \ \delta_{\text{X}_i,\text{X}_j} r_{\text{X}_j}^{(1,1)} s_{\text{X}_i}^{(1,1)} + \left(1-\delta_{\text{X}_i,\text{X}_j}\right) r_{\text{X}_j}^{(0,1)} s_{\text{X}_i}^{(0,1)}\ ,
    \label{eq:w_2}
\end{align}

\noindent being $\delta_{x,y}$ a Kronecker delta.

If the interaction is part of a $3$-edge, the transmission probability can take eight different values, depending on the type of all the three nodes, say $i$, $j$ and $l$. Focusing on the interaction between $i$ and $j$, if the former is susceptible and the other one is infected, we have:
$\lambda r_{\text{X}_j}^{(2,2)} s_{\text{X}_i}^{(2,2)}$ if $\text{X}_i = \text{X}_j = \text{X}_l$; $\lambda r_{\text{X}_j}^{(1,2)} s_{\text{X}_i}^{(1,2)}$ if $\text{X}_i = \text{X}_j \neq \text{X}_l$; $\lambda r_{\text{X}_j}^{(1,2)} s_{\text{X}_i}^{(0,2)}$ if $\text{X}_i \neq \text{X}_j$ and $\text{X}_j = \text{X}_l$; and $\lambda r_{\text{X}_j}^{(0,2)} s_{\text{X}_i}^{(1,2)}$ if $\text{X}_i \neq \text{X}_j$ and $\text{X}_i = \text{X}_l$. Then, the probability $w_i^{(3)}(t)$ that $i$ does not get infected within any of the $3$-edges it takes part to, reads
\begin{align}
    \notag w_i^{(3)}(t) = \prod_{j=1}^N\prod_{l>j}^N A_{i,j,l}^{(3)}&\left[1 - \sigma_j(t)\lambda \tau_{i,j;l}^{(3)} \Delta t\right] \\
    & \times \left[1 - \sigma_l(t)\lambda \tau_{i,l;j}^{(3)} \Delta t\right]\ ,
    \label{eq:no_inf_3}
\end{align}

\noindent where
\begin{align}
    \notag \tau_{i,j;l}^{(3)} \equiv &\ \tau^{(3)}\left(\text{X}_i,\text{X}_j;\text{X}_l\right) \\
    \notag =& \ \delta_{\text{X}_i,\text{X}_j} \\
    \notag &\times \left[\delta_{\text{X}_i,\text{X}_l} r_{\text{X}_j}^{(2,2)} s_{\text{X}_i}^{(2,2)} + \left(1-\delta_{\text{X}_i,\text{X}_l}\right) r_{\text{X}_j}^{(1,2)} s_{\text{X}_i}^{(1,2)}\right] \\
    \notag &+ \left(1-\delta_{\text{X}_i,\text{X}_j}\right) \\
    &\times \left[\delta_{\text{X}_i,\text{X}_l} r_{\text{X}_j}^{(0,2)} s_{\text{X}_i}^{(1,2)} + \left(1-\delta_{\text{X}_i,\text{X}_l}\right) r_{\text{X}_j}^{(1,2)} s_{\text{X}_i}^{(0,2)}\right]\ .
    \label{eq:w_3}
\end{align}

All in all, the probability that a susceptible node $i$ gets infected via at least one of its contacts is
\begin{equation}
    P\left(\sigma_i(t+\Delta t) = 1\vert \sigma_i(t) = 0\right) = 1 - w_i^{(2)}(t)w_i^{(3)}(t)\ ,
    \label{P_inf}
\end{equation}

\noindent while its probability of recovery if infected is simply
\begin{equation}
    P\left(\sigma_i(t+\Delta t) = 0\vert \sigma_i(t) = 1\right) = \mu \Delta t\ .
    \label{P_rec}
\end{equation}

The stochastic process is implemented by drawing a value for $u_i\sim\text{Uniform}(0,1)$ independently for each node $i$, and at each time step. If node $i$ is susceptible, it gets infected if $P\left(\sigma_i(t+\Delta t) = 1\vert \sigma_i(t) = 0\right) > u_i$; if infected, it recovers if $P\left(\sigma_i(t+\Delta t) = 0\vert \sigma_i(t) = 1\right) > u_i$.

The continuous dynamics corresponds to the limit $\Delta t \rightarrow 0$, as only the terms linear in $\Delta t$ survive. In the simulation, we took $\Delta t = 0.05$. Using the quasistationary state (QS) method (with $50$~stored active states and a $25\%$ probability of update)\cite{ferreira2012epidemic}, we let the system run for a transient of $1000$ time steps and then averaged over the last $500$ ones.

\clearpage

\renewcommand\appendixname{}

\renewcommand\theequation{{S\arabic{equation}}}
\renewcommand\thetable{{S\Roman{table}}}
\renewcommand{\suppfigurename}{FIG.}
\renewcommand\thesuppfigure{{S\arabic{suppfigure}}}
\renewcommand\thesection{{S\arabic{section}}}
\renewcommand\thesubsection{{S\arabic{section}.\arabic{subsection}}}

\setcounter{section}{0}
\setcounter{table}{0}
\setcounter{figure}{0}
\setcounter{equation}{0}

\onecolumngrid

\section*{Supplemental Material}

\section{Social contagion model}

We show here the equilibrium adoption probabilities for the social contagion model introduced in Sec.~IIIA of the main text. Fig.~\ref{fig_supp_1} reports them for the two types as a function of the composition of a group for different values of the A-type's adoption rate, $b_\text{A}$ (the others rates are fixed as in the main text, $c_\text{A}=c_\text{N}=0.05$ and $b_\text{N}=c_\text{N}/20$). One can clearly observe the nonlinear dependence on the composition. The non-linearity increases with $b_\text{A}$ and tends to disappear only for $b_\text{A}$ approaching $c_\text{A}$.

\begin{suppfigure}[h]
    \centering
    \includegraphics[width = 1\linewidth]{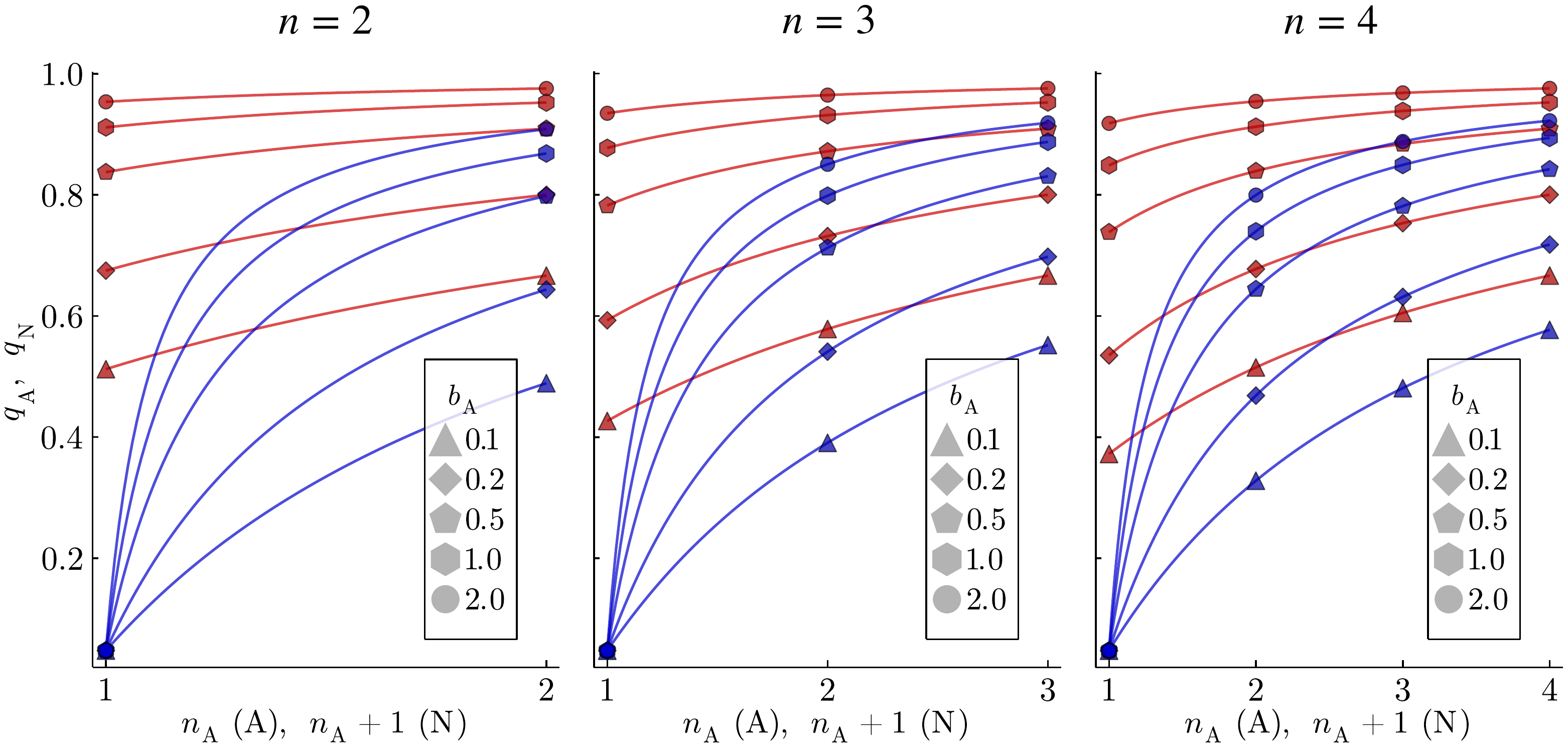}
    \caption{Equilibrium adoption probability for A-type individuals, $q_{\text{A}}$ (red points), and for N-type ones, $q_{\text{N}}$ (blue points), against the number of A-type's in the group (from $1$ to $n$ for an A-type's, from $0$ to $n-1$ for a N-type's). Different symbols denote different adoption rates for A-type's, $b_\text{A}$, while the other rates are fixed as $c_\text{A}=c_\text{N}=0.05$ and $b_\text{N}=c_\text{N}/20$. In the main text we used $b_\text{A}=1.0$ (hexagons). Left plot is for $2$-edges ($n=2$), middle plot for $3$-edges ($n=3$) and right plot for $4$-edges ($n=4$). To better appreciate the nonlinear dependence on $n_\text{A}$, we also depicts through solid lines the equilibrium solutions as if $n_\text{A}$ was a continuous variable.}
    \label{fig_supp_1}
\end{suppfigure}

\section{Endemic state for the two binary models}

In this section we show the results for the endemic equilibrium in the two limits of easy and hard adaptation, useful to properly interpret the results obtained for intermediate scenarios like the ones entailed by the social contagion model. We first vary the $2$- and $3$-degree for a fixed overall degree $k = k^{(2)} + 2k^{(3)}$, for given values of the mixing parameters ($\alpha_2$, $\alpha_3$ and $\beta_{3,1}$). Then we vary the pairwise assortativity distribution between pairs ($\alpha_2$) and triads ($\alpha_3$) for a fixed average pairwise assortativity $\alpha = \left(\alpha_2 k^{(2)} + 2\alpha_3 k^{(3)}\right)/\left(k^{(2)} + 2k^{(3)}\right)$, assumed $k_\text{A}^{(n)}=k_\text{N}^{(n)}=k^{(n)}$, $n=2,3$. In both cases we use $\varepsilon_\text{out} = 0.9$ and $\varepsilon_\text{in} = 0.5$. Results are shown in Figs.~\ref{fig_supp_2}-\ref{fig_supp_5}. The results found in the hard adaptation limit when varying the degree distribution among group sizes are analysed in Sec.~S2.1, while all the others are already discussed in the main text.

\subsection{Varying the degree distribution among group sizes: hard adaptation}

Here we provide a detailed analysis of the results obtained in the limit of hard adaptation when varying the distribution of the degree between the two group sizes (see Fig.~\ref{fig_supp_5}). In this behavioral scenario, $I_\text{A}$ is the main driver of the observed phenomenology. For $\alpha_2 = \alpha_3$ (recall $\beta_{3,1}=1$), $I_\text{A}$ decreases with $k^{(2)}$, for the A-N interactions in which the N-type individual is also an adopter becomes more frequent. Since the N-type individuals self-protect more often, also $I_\text{N}$ decreases, but only slightly, as the epidemic pressure from the type A is small (unless efficacy is low in both ways). For $\alpha_2 < \alpha_3$, the effect on $I_\text{N}$ produced by the more frequent N-N interactions, as implied by a raised $k^{(2)}$, overcomes the little benefit coming from the protection in the fewer A-N interactions, hence $I_\text{N}$ increases (similarly to $R$). This growth is faster for high values of $\rho$, for the $\{\text{A},\text{N},\text{N}\}$ triads become scarce and only few protected A-N interactions are gained interacting within pairs. Moreover, for high enough $\rho$, the growing pressure coming from type N is able to compensate the benefit for type A entailed by a lower mixing with type N. When the efficacy is very high (and hence mixing is important), $I_\text{A}$ can even show a non-monotonous shape, with a maximum at small values of $k^{(2)}$; otherwise it always decreases with the latter. On the other hand, for low $\rho$ (meaning A-type's mainly find themselves in the highly infectious $\{\text{A},\text{N},\text{N}\}$ configuration when interacting in triads) and poor protection in both ways, the relative decrease of $I_\text{A}$ is sufficiently rapid to even induce a slight decrease of $I_\text{N}$.

Considering $\alpha_2 > \alpha_3$, $I_\text{N}$ (like $R$) decreases with $k^{(2)}$ as a consequence of the N-N interactions becoming sparser and, secondarily, of the A-N interactions being more often bilaterally protected. The decrease is faster when $\rho$ is high as the many A-type individuals can better decrease the epidemic pressure among the N-type's. The increased rate at which A-N interactions are fully protected generally makes $I_\text{A}$ decrease as well. An exception to this exists for low values of $\rho$ when the in-going and out-going efficacy are, respectively, high and low. In such situation, on one hand, since the efficacy is high in at least one way (in-going), the two types have sensibly different epidemic pressures, therefore the mixing matters and for type A is convenient to reduce the contacts with type N; on the other, the transmission coming from type N is only poorly stopped ($\varepsilon_\text{out}$ low), hence the more frequent adoption from type N, as entailed by a larger $k^{(2)}$, benefits type A only marginally. Altogether, type A finds it convenient to mix less with type N, as implied by a smaller $k^{(2)}$, but only if $\rho$ is not too high, otherwise the rapid decrease of $I_\text{N}$ with $k^{(2)}$ drives $I_\text{A}$ to decrease as well.

\begin{suppfigure}
    \centering
    \includegraphics[width = 1\linewidth]{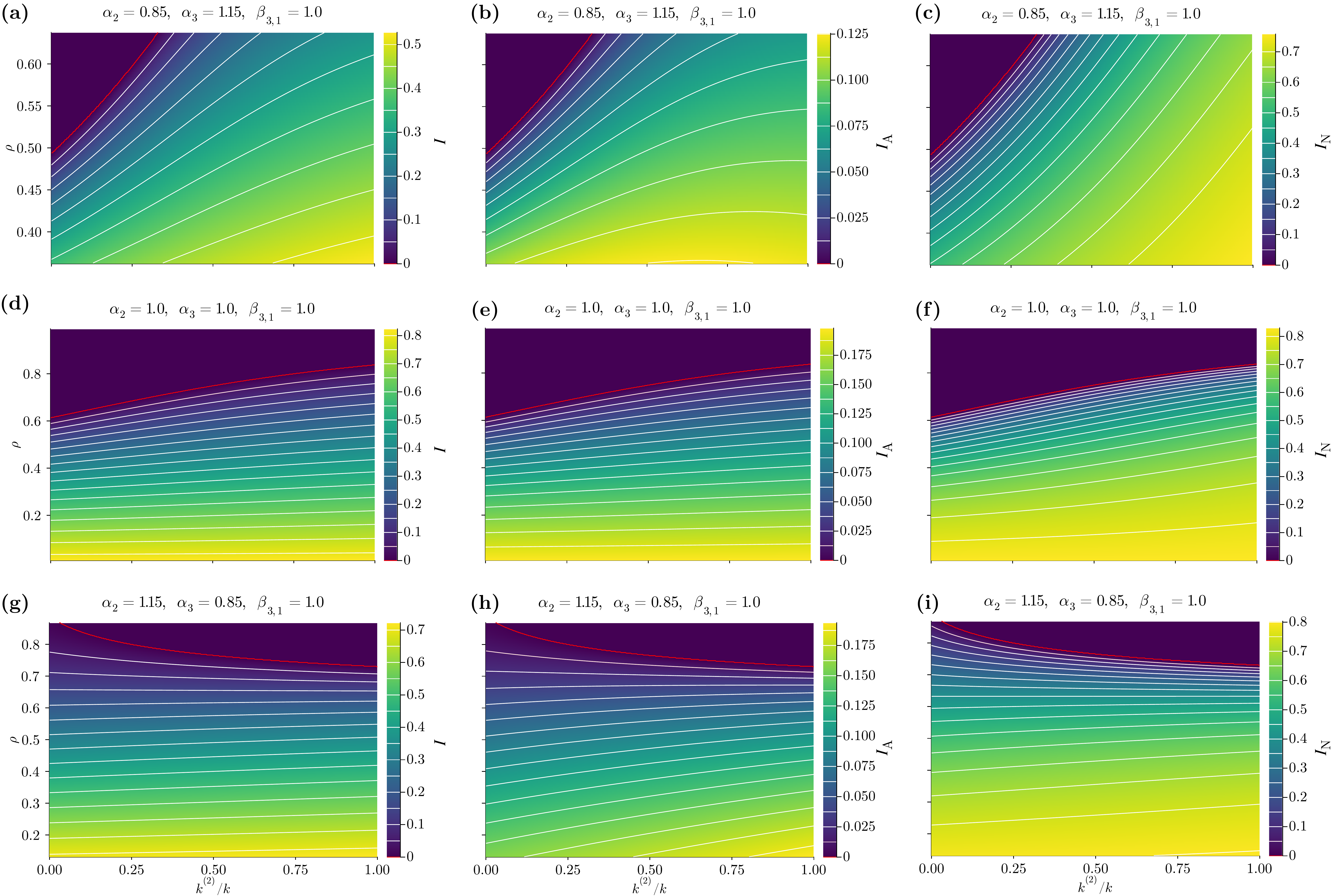}    \caption{{\bf Easy adaptation.} Equilibrium endemic state as a function of $k^{(2)}/k$ and $\rho$ for $\beta_{3,1}=1.0$, $R_0 = k = 6$, and {\bf (a--c)} $\alpha_2 = 0.85$ and $\alpha_3 = 1.15$, {\bf (d--f)} $\alpha_2 = \alpha_3 = 1.0$, {\bf (g--i)} $\alpha_2 = 1.15$ and $\alpha_3 = 0.85$. {\bf (a, d, g)} Overall prevalence, $I$, {\bf (b, e, h)} prevalence for type A, $I_\text{A}$, and {\bf (c, f, i)} prevalence for type N, $I_\text{N}$. White curves report the levels indicated in the respective colorbar. The red one indicates the critical curve $R=1$, above which the disease-free state is stable.}
    \label{fig_supp_3}
\end{suppfigure}

\begin{suppfigure}
    \centering
    \includegraphics[width = 1\linewidth]{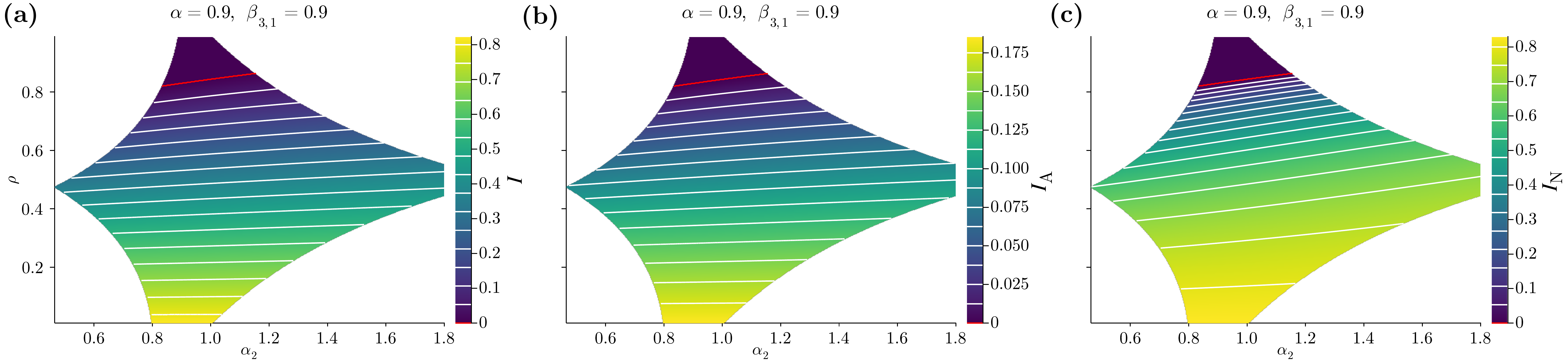}
    \caption{{\bf Easy adaptation.} Equilibrium endemic state as a function of $\alpha_2$ and $\rho$ for $\alpha=0.9$, $\beta_{3,1}=0.9$, $k^{(2)}=2k^{(3)}$ and $R_0 = k = 6$. {\bf (a)} Overall prevalence, $I$, {\bf (b)} prevalence for type A, $I_\text{A}$, and {\bf (c)} prevalence for type N, $I_\text{N}$. White curves report the levels indicated in the respective colorbar. The red one indicates the critical curve $R=1$, above which the disease-free state is stable.}
    \label{fig_supp_2}
\end{suppfigure}

\begin{suppfigure}
    \centering
    \includegraphics[width = 1\linewidth]{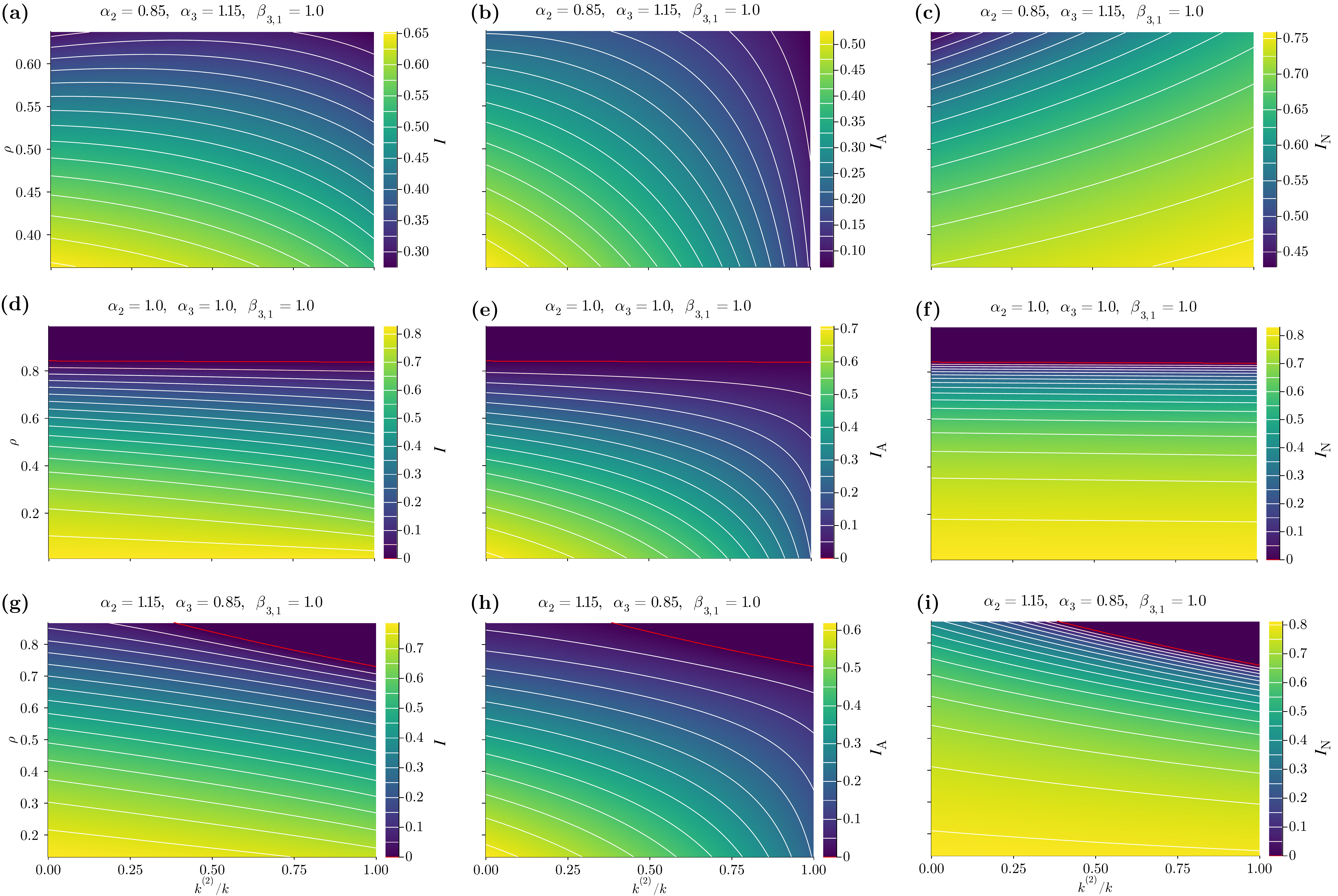}    \caption{{\bf Hard adaptation.} Equilibrium endemic state as a function of $k^{(2)}/k$ and $\rho$ for $\beta_{3,1}=1.0$, $R_0 = k = 6$, and {\bf (a--c)} $\alpha_2 = 0.85$ and $\alpha_3 = 1.15$, {\bf (d--f)} $\alpha_2 = \alpha_3 = 1.0$, {\bf (g--i)} $\alpha_2 = 1.15$ and $\alpha_3 = 0.85$. {\bf (a, d, g)} Overall prevalence, $I$, {\bf (b, e, h)} prevalence for type A, $I_\text{A}$, and {\bf (c, f, i)} prevalence for type N, $I_\text{N}$. White curves report the levels indicated in the respective colorbar. The red one indicates the critical curve $R=1$, above which the disease-free state is stable.}
    \label{fig_supp_5}
\end{suppfigure}

\begin{suppfigure}
    \centering
    \includegraphics[width = 1\linewidth]{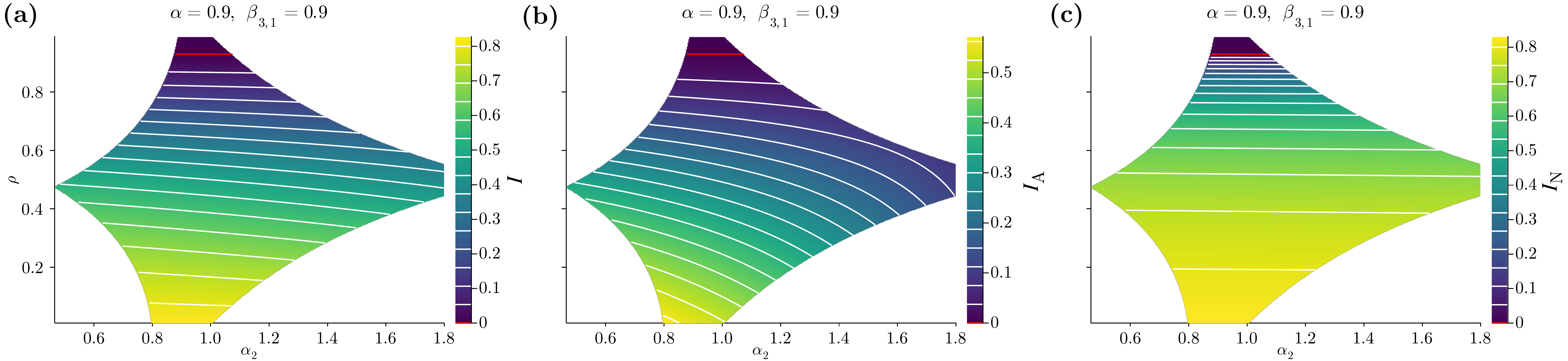}
    \caption{{\bf Hard adaptation.} Equilibrium endemic state as a function of $\alpha_2$ and $\rho$ for $\alpha=0.9$, $\beta_{3,1}=0.9$, $k^{(2)}=2k^{(3)}$ and $R_0 = k = 6$. {\bf (a)} Overall prevalence, $I$, {\bf (b)} prevalence for type A, $I_\text{A}$, and {\bf (c)} prevalence for type N, $I_\text{N}$. White curves report the levels indicated in the respective colorbar. The red one indicates the critical curve $R=1$, above which the disease-free state is stable.}
    \label{fig_supp_4}
\end{suppfigure}

\section{Varying the type-assortativity distribution among group sizes: intermediate, social contagion scenario}

We here focus on the effect of distributing the assortativity differently between $2$- and $3$-edges considering the intermediate scenario used in the main text and derived from the social contagion model of Sec.~IIIA.

\begin{suppfigure}
    \centering
    \includegraphics[width = 0.95\linewidth]{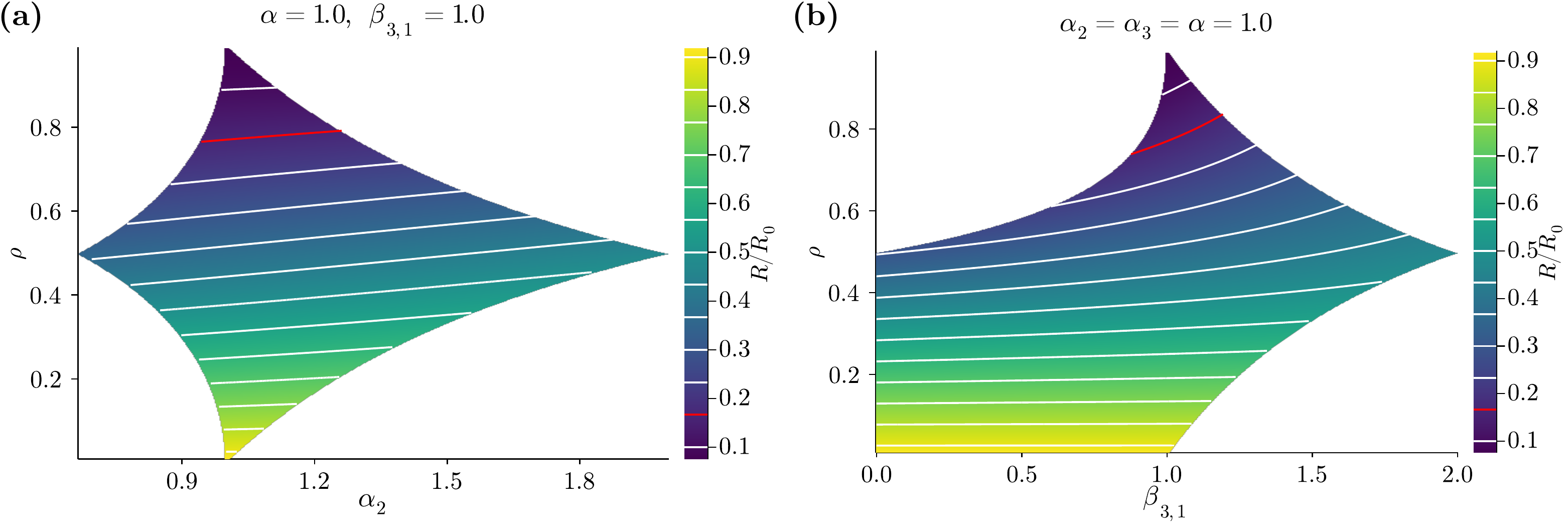}
    \caption{{\bf Intermediate scenario.} Normalized reproduction number, $R/R_0=k_\text{eff}/k$, as a function of $\rho$ and of {\bf (a)} $\alpha_2$ ($\alpha=1.0$, $\beta_{3,1}=1.0$), and {\bf (b)} $\beta_{3,1}$ ($\alpha_2=\alpha_3=\alpha=1.0$). We fixed $k^{(2)}=2k^{(3)}$ and $R_0 = k = 6$. White curves report the levels indicated in the colorbar. The red one indicates the curve $R=1$ ($R/R_0=1/6$), above which the disease-free state is stable.
    }
    \label{fig_supp_6}
\end{suppfigure}

\subsection{Reproduction number}
\label{sec:results_assor_R}

As Fig.~\ref{fig_supp_6}(a) shows, the basic reproduction number, $R$, increases (decreases) with $\alpha_2$ ($\alpha_3$), whatever is the proportion $\rho$ of A-type individuals in the population. In other words, for a given overall assortativity, the minimal (maximal) (dis)assortativity within triads is the optimal mixing scheme to minimize the early growth rate of an outbreak. This is due to the strong behavioral adaptation of N-type individuals under the presence of just one A-type's in the group. Indeed, since $q_\text{N}$ jumps from around $0.05$ to $0.80$ when going from a $\{\text{N},\text{N},\text{N}\}$ triad to a $\{\text{A},\text{N},\text{N}\}$ one, a single A-type suffices to notably reduce the transmission probability of all the direct interactions in the triad, including the N-N one. Minimizing the assortativity within triads means maximizing the rate at which pairs of N-type's gather with an A-type (see Eqs.~(14) and (18) of the main text), and therefore the rate at which N-N direct interactions are indirectly weakened (the A-N ones are anyway weak due to the A-type). This intuition can be rigorously proven considering the binary scenario of \lq easy adaptation\rq\ (see Appendix~C2 of the main text), which is a proxy for the one we are using here.

The triads $\{\text{A},\text{N},\text{N}\}$ are the responsible for the dependence of $R$ on the distribution of assortativity, as they are the only ones within which the behavioral outcome changes passing from one limit scenario to the other. Accordingly, in this intermediate scenario, reducing the frequency of $\{\text{A},\text{N},\text{N}\}$ triads by increasing $\beta_{3,1}$ (see Eq.~(15) of the main text), makes $R$ grow. We report this in Fig.~\ref{fig_supp_6}(b). What the latter also shows is that the pairwise information alone (i.e., knowing $\alpha_2$ and $\alpha_3$) is generally not sufficient to guarantee an accurate description of the system.

\subsection{Prevalence}
\label{sec:results_assor_prev}

Looking now at the levels of prevalence at equilibrium, we can appreciate how differently the two types are affected by the assortativity distribution. In Fig.~\ref{fig_supp_7} we assume a moderate overall assortativity by taking $\alpha=0.9$, and we fix $\beta_{3,1}=0.9$. The phenomenology is qualitatively unaltered by choosing other values for these parameters, whose effect is mainly changing the ranges within which $\alpha_2$ and $\alpha_3$ can vary (in particular, $\alpha\neq 1$ shifts the ranges, while $\beta_{3,1}\neq 1$ makes them asymmetric with respect to the exchange $\rho \leftrightarrow 1-\rho$).

\begin{suppfigure}
    \centering
    \includegraphics[width = 1.\linewidth]{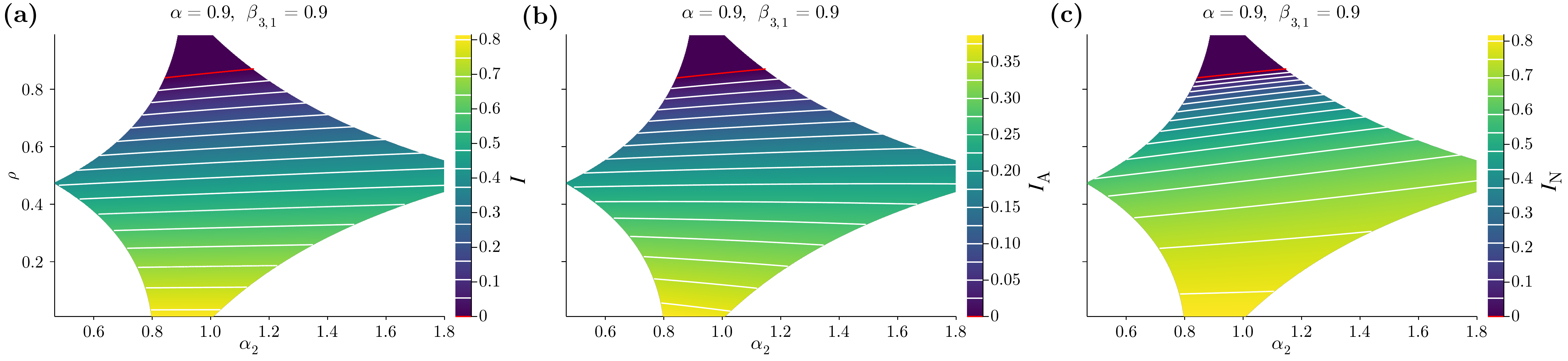}
    \caption{{\bf Intermediate scenario.} Equilibrium endemic state as a function of $\alpha_2$ and $\rho$ for $\alpha=0.9$, $\beta_{3,1}=0.9$, $k^{(2)}=2k^{(3)}$ and $R_0 = k = 6$. {\bf (a)} Overall prevalence, $I$, {\bf (b)} prevalence for type A, $I_\text{A}$, and {\bf (c)} prevalence for type N, $I_\text{N}$. White curves report the levels indicated in the respective colorbar. The red one indicates the curve $R=1$, above which the disease-free state is stable.}
    \label{fig_supp_7}
\end{suppfigure}

We observe that, while $I_\text{N}$ behaves like $R$, $I_\text{A}$ may not in general (see the discussion in the main text and the analysis in Appendix~D). For instance, from Fig.~\ref{fig_supp_7} we see that $I_\text{A}$ increases with $\alpha_2$ as $I_\text{N}$ and $R$ do, but only if $\rho$ is high enough, otherwise it decreases. The intermediate scenario considered is closer to the easy adoption one, which is why we see a wide overlap between the two. Nonetheless, for small $\rho$, we know that $I_\text{A}$ is barely affected by $\alpha_2$ in easy adaptation, whereas the effect is strong in hard adaptation. The fact that, differently than in the easy adaptation scenario, $q_\text{N}$ in a $\{\text{A},\text{N},\text{N}\}$ triad is not as high as in a $\{\text{A},\text{N}\}$ pair (or in a $\{\text{A},\text{A},\text{N}\}$ triad), accounts for the decrease observed for $I_\text{A}$. Consistently, the decrease is small compared to the one found in the hard adaptation scenario (see Fig.~S4).

All in all, for any intermediate scenario, we can generally say that lowering the assortativity within triads makes the endemic prevalence decrease (or, at least, not increase) for type N, and increase or decrease for type A depending on whether the share of A-type individuals, $\rho$, stays below or above a certain value.

Lastly, as already seen for the basic reproduction number, increasing $\beta_{3,1}$ has the same qualitative effect of increasing $\alpha_2$.


\providecommand{\noopsort}[1]{}\providecommand{\singleletter}[1]{#1}%

\end{document}